\documentclass[11pt]{article}
\usepackage[utf8]{inputenc}
\usepackage{graphicx} 
\usepackage{xcolor}
\usepackage[top = 1 in, bottom = 1 in , left= 1 in, right = 1 in]{geometry}
\usepackage[numbers,sort&compress]{natbib}
\usepackage{todonotes}
\usepackage{soul}
\setstcolor{red}
\usepackage{array,makecell,booktabs,tabularx}
\usepackage{float}
\usepackage{amsmath,amssymb,amsfonts,bm}
\usepackage{empheq}
\usepackage{subfigure}
\usepackage{array,threeparttable}
\usepackage{multirow}
\usepackage{booktabs}
\usepackage{siunitx}
\usepackage{authblk}

\providecommand{\keywords}[1]
{
  \small	
  \textbf{\textbf{Keywords: }} #1
}

\title{Natural frequency estimation using complex-frequency excitations}
\author[1]{Wenbo Li}
\author[1]{Raj Kumar Pal \thanks{Corresponding author. Email: rkpal@tamu.edu}}
\affil[1]{J. Mike Walker '66 Department of Mechanical Engineering, Texas A\&M University, College Station, TX 77843, USA}

\begin{document}
\maketitle

\begin{abstract}
%% Text of abstract
%Complex-frequency excitation, as an emerging dynamical approach, holds significant promise for applications such as structural health monitoring and advanced sensor design. However, its distinctive post-processing procedure amplifies the inherent noise of the system, leading to non-negligible effects. To date, the sensitivity of complex-frequency excitation to noise remains largely unexplored. We investigate the dependence of the Fisher information on the excitation and system parameters for a second-order underdamped linear time-invariant system driven by the complex-frequency excitation, under both Gaussian white and colored noise. Closed-form expressions are derived to characterize the relationship between the Fisher information and the decay factor of the complex-frequency excitation, demonstrating that a substantial enhancement in Fisher information can be achieved under appropriate parameter selections. The theoretical predictions are verified through numerical simulations. This work reveals the superiority of complex-frequency excitation over conventional approaches and establishes a theoretical foundation for the design of advanced sensors and more efficient nondestructive evaluation methods, from an information-theoretic perspective. 
Complex frequency excitations, oscillating signals whose amplitude decreases exponentially in time, have recently been demonstrated to significantly increase the effective quality factor of mechanical resonators. In this work, we investigate the accuracy of natural frequency estimation in mechanical systems under noise using such excitations. The analysis is performed on an underdamped linear time-invariant single-degree-of-freedom spring-mass-damper system. We employ tools from information theory, namely Fisher information, to systematically quantify the sensitivity of complex-frequency excitation to measurement noise. Explicit closed-form expressions are derived relating Fisher information to excitation and system parameters under both Gaussian white and colored noise. The theoretical predictions are verified through Monte Carlo numerical simulations. 
The results indicate that appropriate selection of excitation parameters can significantly enhance the Fisher information, leading to improved estimation accuracy under complex-frequency excitations compared with conventional harmonic excitations.
Experimental results demonstrate the advantages of complex-frequency excitation in terms of both accuracy and robustness of natural-frequency estimation. These findings establish a foundation for the development of high-performance sensors and novel nondestructive evaluation methods.
\end{abstract}


\section{Introduction}
% outline of introduction:
% P1: measuring freq. is essential, existing methods and their limits
% P2: complex freq provide new physics
% P3: complex freq. provides a new way to significantly enhance Q_eff
% P4: limitations and open questions in our work
% P5: paper outline

Measuring natural frequencies is essential for vibration control and structural health monitoring of diverse mechanical, aerospace, and civil engineering structures \cite{ewins2009modal,hou2021review,sepahvand2017stochastic}. A wide range of methods have been developed over the past few decades for natural frequency estimation from vibration data~\cite{peeters2003comparative,devriendt2008identification}. They fall broadly into two categories: time-domain and frequency-domain based. In time-domain methods, the free-decay response to an impulse is represented as a superposition of complex exponential terms and the modal parameters are identified through fitting. Representative approaches include Prony-type exponential fitting method \cite{hauer2002application}, the Eigensystem Realization Algorithm \cite{juang1985eigensystem,caicedo2011practical}, and related subspace identification frameworks \cite{overschee1996subspace,peeters1999reference,sun2025analytical}. Frequency-domain methods, by contrast, use frequency response functions (FRFs) obtained under harmonic or broadband excitation, from which modal parameters are extracted by parametric curve fitting of the transfer function. Typical approaches employ rational-function or polynomial fitting forms, such as the PolyMAX estimator \cite{peeters2004polymax}, other frequency-domain modal-fitting \cite{jacquelin2015polynomial,maia2001modal} and data-driven methods~\cite{padil2020non,liu2023data}. In practical applications, the presence of noise limits the accuracy and robustness of frequency estimation, particularly when resonance peaks can no longer be reliably identified from the FRF under strong noise conditions~\cite{kara2023two,peeters2004polymax,worden2007application, garcia2020automated,klapuri2004multiple,kumar2025critical,rahman2021orthogonal}. To overcome this obstacle, here we consider a wider class of excitations beyond harmonics and impulses, called complex frequency excitations, with the goal of accurate natural frequency estimation even under high-intensity noise.

Complex-frequencies are a generalization of real frequencies that lead to harmonic excitations $e^{i \omega t}$, where the frequency $\omega = \omega_{real} + i\Omega$ is a complex number. Their signals comprise an oscillatory component $e^{i\omega_{real} t}$ whose amplitude decays exponentially with time $e^{-\Omega t}$. Recent studies with these excitations have shown novel dynamics phenomena in multiple physical domains, including plasmonics, optics and elastodynamics \cite{guan2024compensating,an2025complex,trivedi2025selective,wu2021vibration,sweeney2020theory,baranov2017coherent}. For instance, in plasmonics, complex-frequency excitation has been successfully exploited to overcome dissipation-induced limits on scattering and imaging resolution \cite{kim2022beyond,guan2023overcoming}. In optics, Alu and coworkers predicted directional scattering enhanced over two orders in magnitude compared to harmonic excitations \cite{kim2022beyond}. Similarly, in elastodynamics, it has enabled complete energy trapping of elastic waves in plates for small times~\cite{rasmussen2023lossless}. 
The physical intuition behind these novel phenomena is that the rate at which energy is stored and released from the system is distinct compared to harmonic excitations, \textit{i.e.,} the system has a virtual gain or loss~\cite{kim2025complex}. Such excitations open avenues for novel system identification and NDE techniques that can surpass the limits of conventional harmonic ones.

Recently, the authors have investigated the dynamics of a damped cantilever beam under complex frequency excitations~\cite{li2025effective}. They demonstrated a significant increase ($\sim 250\times$) in its effective quality factor compared to harmonic excitations. The quality factor of a structure is a key parameter that governs the resolution~\cite{miller2018effective} of natural frequency estimation. Conventional approaches to boost quality factor, for instance in micro- and nano-scale sensors, primarily rely on redesigning the device geometry or implementing real-time closed-loop control of the resonator \cite{dania2024ultrahigh,mahashabde2020fast,metzger2008self,taheri2017mutual}. These strategies are often costly and complex to implement, and they typically provide only limited improvements in the effective quality factor (about 5–20 fold) \cite{miller2018effective}. In contrast, complex-frequency excitations use a relatively simple open-loop scheme without modifying the structure or implementing feedback to attain performance improvements. Exploiting such excitations thus may have distinct advantages over traditional methods.

The amplitude of complex-frequency excitations ($e^{-\Omega t}\cos \omega_{real} t$) can either increase or decrease in time depending on the sign of the imaginary frequency component $\Omega$. The excitations relevant for frequency estimation~\cite{li2025effective} have  amplitudes decaying exponentially in time. In an ideal noiseless environment, such excitations can lead to extremely high effective quality factors limited by only the sensor and actuator resolution even in damped structures. In practice, however, the presence of noise implies that the rapidly decaying excitation amplitude will cause the corresponding system output to be overwhelmed by noise. This effect of noise imposes stringent constraints on the applied complex-frequency excitation, such as the allowable excitation duration and the amplitude decay rate, both of which affect the output signal-to-noise ratio (SNR) over time. 
For effective implementation of complex-frequency excitations for frequency estimation, it is thus essential to understand their performance under noisy conditions. 

Here, we employ tools from information theory to analyze a model system comprising an underdamped linear time-invariant spring-mass-damper with Gaussian noise. To carefully identify the effect of various parameters, we derive expressions for the Fisher information ratio between complex-frequency and conventional harmonic excitations. The accompanying simulations and analysis with experimental data-set reveal that information gain depends on both the excitation and system parameters. 
The remainder of this paper is structured as follows. In Sec. \ref{Theory}, we first introduce our natural frequency estimation framework. Then, we derive the Fisher information ratio of a system with Gaussian white and colored noise. In Sec. \ref{Numerical simulations}, the theoretical predictions are verified through Monte Carlo simulations. In Sec. \ref{Experimental validations}, we add noise to an experimental dataset to compare the natural frequency estimation accuracy with harmonic versus complex-frequency excitations. Finally, discussion and conclusions are presented in Sec. \ref{Conclusions}.

\section{Theory}
\label{Theory}

We first give an overview of the natural frequency estimation method using complex-frequency excitations. The key observation is that since excitation amplitude decays exponentially in time, the input signal and the resulting output are dominated by noise after some time. This decay of the signals in time makes it essential to carefully analyze the role of each parameter (decay rate, amplitude, excitation time) on the accuracy of frequency estimation. The Fisher information is employed to analyze the performance of the method on a system under Gaussian white noise and Gaussian colored noise. Explicit expressions for the Fisher-information gain ratio when compared with harmonic excitations are derived under equal-initial-amplitude and equal-energy conditions.

It should be noted that, in this section, frequency estimation refers to the estimation of the frequency involved in the spectral reconstruction of the noisy mapped output within the complex-frequency excitation framework. The purpose of this frequency estimation is to obtain a more accurate and robust amplitude-frequency response function, and thereby a more accurate and robust estimate of the natural frequency. It does not refer to the direct estimation of the system natural frequency itself.

\subsection{Background: frequency estimation with complex-frequency excitations}
%The complex-frequency excitation framework achieves control of effective energy gain or loss by applying complex-frequency modulation to the original excitation and inverse complex-frequency processing to the actual output. 

\begin{figure}[h!]%fig1
	\makeatletter
	\renewcommand{\fnum@figure}{Fig. \thefigure.\@gobble}
	\makeatother
	\centering
	\includegraphics[scale=0.47]{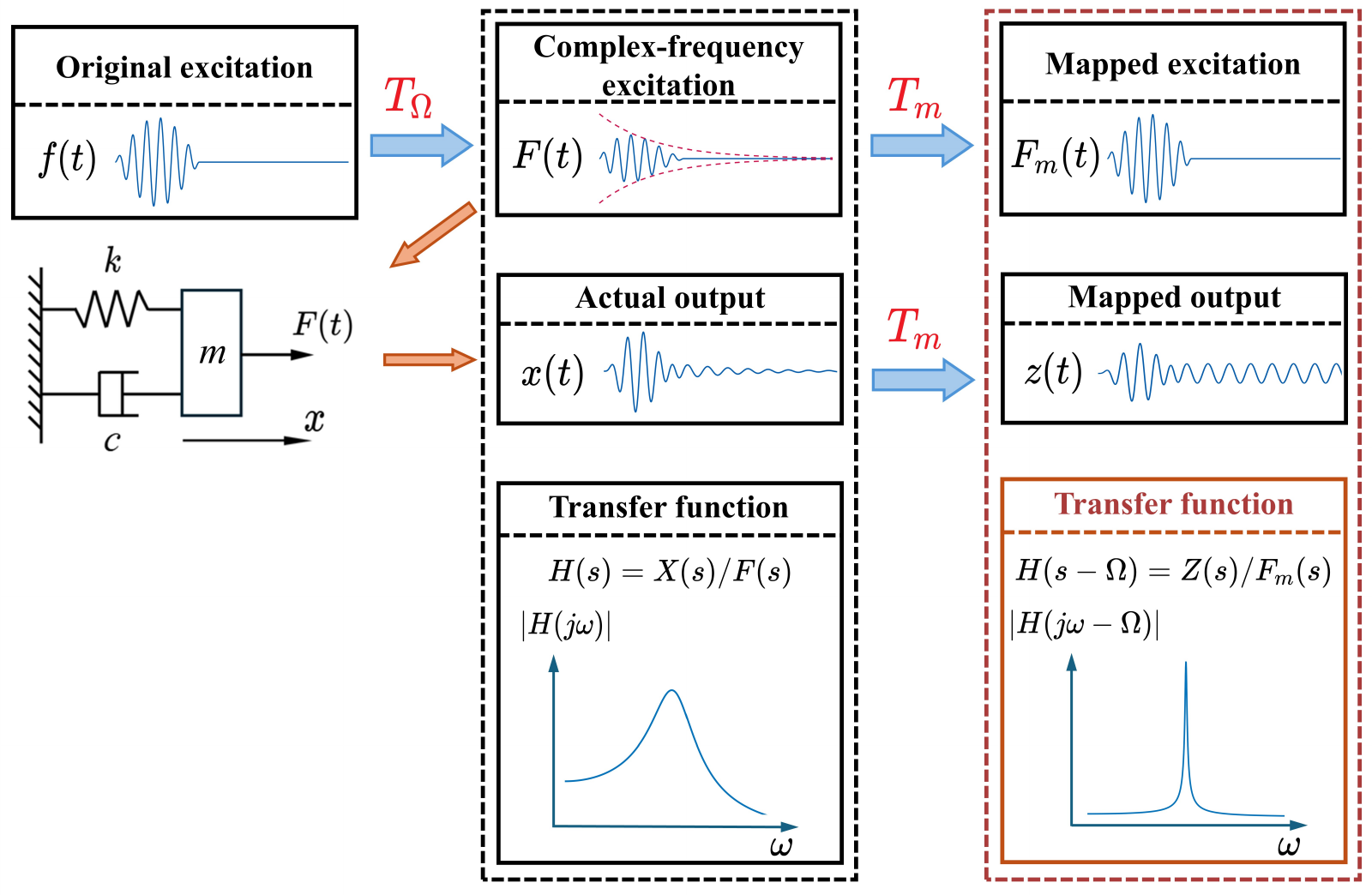}
	\caption{Basic principle of the complex-frequency excitation framework. An operator $T_{\Omega}$ is applied to the original excitation $f(t)$ to generate $F(t)$, which is the actual excitation applied to the mass–spring–damper system. The excitation force $F(t)$ and resulting displacement $x(t)$ are then subject to the mapping $T_m = T_{\Omega}^{-1}$ to obtain the mapped excitation $F_m(t)$ and mapped output $z(t)$. $F_m(t)$ and $z(t)$ are used to determine the modified transfer function for frequency estimation.}
	\label{fig2-1}
\end{figure}

%The complex-frequency excitation %\textbf{ (or should it be nat. freq. estimation framework??)} 
%framework is formulated for a second-order underdamped linear time-invariant system, and its basic principle is illustrated using the single-degree-of-freedom mass–spring–damper system shown in Fig. 1. As depicted in Fig. 1, the original external drive is defined as the original excitation $f(t)$. By applying complex-frequency processing $T_{\Omega}$ to $f(t)$, we obtain a new excitation whose amplitude varies exponentially in time, which is then used as the actual input to the system. This actual input is also referred to as the complex-frequency excitation $F(t)$. Similarly, by applying the mapping $T_{\Omega}^{-1}$, to $F(t)$, the mapped excitation $F_m(t)$ is obtained. We assume that the system operates in a quasi-steady state under $F(t)$. The displacement of the mass is defined as the actual output $x(t)$, and the mapping $T_{\Omega}^{-1}$ is applied to $x(t)$ to obtain the final output of the framework, i.e., the mapped output $z(t)$.
Natural frequency estimation accuracy depends on both external (noise) and internal (damping) factors. In particular, a higher damping causes the frequency response function (FRF) near resonances to be broader, thereby lowering resolution. FRF corresponds to the steady state response under harmonic excitations. Here, we show how examining the response of an appropriate transformed variable under  complex-frequency excitations can yield an FRF that is significantly narrower near resonance, approaching the behavior of an undamped system. 
Figure~\ref{fig2-1} displays an overview of the framework presented with a single-degree-of-freedom mass-spring-damper system. It is applicable to any second-order underdamped linear time-invariant system. The idea is to examine the response of an equivalent system that is obtained by applying appropriate mappings. To this end, we define two operators: a complex frequency operator $T_{\Omega}$ and a mapping $T_m$. Let $g(t)$ be an arbitrary function of time $t$, then these operators are defined as: 
\begin{equation} \label{eq1-2}
	T_{\Omega}[g(t)]=e^{-\Omega t}g(t), 
    \qquad 
    	T_m[g(t)] = T_{\Omega}^{-1}[g(t)]=e^{\Omega t}g(t).
\end{equation}
where $\Omega$ is the imaginary component of a complex-frequency. In this work, we set the mapped excitation to be the inverse of the complex-frequency operator, however in general, $T_m$ can be independent of $T_{\Omega}$. 

Our framework has two steps. We start by considering a short time excitation signal $f(t)$ that is narrow band centered around a frequency $\omega_0$, for example a windowed tone burst. In conventional methods, one can apply this signal to get the transfer function or frequency response at $\omega_0$. Here, $f(t)$ is referred to as the original excitation. Our first step is to apply the complex-frequency operator $T_{\Omega}$ to $f(t)$. The resulting new excitation is  $F(t)=T_{\Omega}[f(t)]$, whose amplitude varies exponentially in time. This signal $F(t)$ is the actual excitation applied to the physical system and is referred to as the complex-frequency excitation. The corresponding system response to $F(t)$ is denoted by $x(t)$, which we call the actual output. We make two remarks here. The first is that $F(t)$ has a broadband in frequency in contrast to the original excitation $f(t)$. The second is that the ratio of output and input signals' Fourier transforms $X(\omega)/F(\omega)$ will have a broad peak near resonance due to damping, see second vertical panel in Fig.~\ref{fig2-1}. 

The second step is to apply the mapping $T_m = T^{-1}_{\Omega}$ to the applied force $F(t)$  and measured displacement $x(t)$, which yields the mapped excitation $F_m(t)=T_{\Omega}^{-1}[F(t)]$ and mapped output $z(t)=T_{\Omega}^{-1}[x(t)]$. Physically, this mapping corresponds to a change of variable or transformation $z(t) = e^{\Omega t} x(t)$. Under this transformation, the governing equation $\ddot{x} + 2 \zeta \omega_n \dot{x} + \omega_n^2 x= F(t)/m$ now becomes
\begin{equation*}
    \ddot{z} + 2 (\zeta \omega_n - \Omega) \dot{z} + (\omega_n^2 - \Omega^2)z = F_m(t)/m . 
\end{equation*}
Here, $m$, $\omega_n$ and $\zeta$  are the mass, natural frequency and damping ratio of the actual system. Note that when $\Omega \approx \zeta \omega_n$, the damping of this mapped system is close to zero and its behavior approaches that of an undamped system with an extremely sharp FRF near resonance, see third vertical panel in Fig.~\ref{fig2-1}. Note that $F(t)$ and $x(t)$ are the actual input and output of the physical system, while the mappings are done to analyze the dynamic response of this transformed variable $z(t)$. Accordingly, $F_m(t)$ and $z(t)$ are used to construct the amplitude-frequency response function to determine the natural frequency. The frequency corresponding to the peak of this function is the damped natural frequency $\omega_d = \omega_n\sqrt{1 - \zeta^2}$ of the system.

Next, we provide a detailed description of the mathematical formulation in the complex-frequency excitation framework. The transfer function of the actual physical system is 
\begin{equation} \label{eq1-1}
	H(s) = \frac{\omega_n^2}{s^2+2\zeta\omega_ns+\omega_n^2}
    = \dfrac{i \omega_n}{2 \sqrt{1-\zeta^2}} 
    \left[ \dfrac{1}{s + \zeta \omega_n + i \sqrt{1-\zeta^2}\omega_n} - 
    \dfrac{1}{s + \zeta \omega_n - i \sqrt{1-\zeta^2}\omega_n}
    \right] .
\end{equation}
Let $f(s)$ be the Laplace spectrum of the original excitation $f(t)$. Then, the Laplace spectrum of the complex-frequency excitation $F(t)=T_{\Omega}[f(t)]$ is 
$F(s)=f(s+\Omega)$. The mapped excitation $F_m(t)$ is 	$F_m(t)=T_{\Omega}^{-1}[F(t)]=f(t)$ and its Laplace spectrum is $F_m(s)=f(s)$.
It is worth noting that although $f(t)$ and $F_m(t)$ are numerically equivalent, they carry distinct physical meanings and have different implications on frequency estimation accuracy. $f(t)$ denotes the original excitation that would be applied to the system to get its conventional FRF, \textit{i.e.,} the FRF under real-valued frequencies. 
In contrast, $F_m(t)$ represents the mapping result of the actual complex-frequency excitation applied to the structure. It is introduced to construct the modified transfer function. As we will show later in this section, the mapping $T_m$ itself does not contribute to any Fisher information gain. Instead, the distinct temporal energy distribution of the complex-frequency excitation $F(t)$ is the key factor that modifies the system dynamics compared to the original excitation $f(t)$. In particular, it is the complex-frequency operator from $f(t)$ to $F(t)$ that gives rise to the Fisher-information gain of the system; this point will be discussed in detail later in this section.

The Laplace spectrum $X(s)$ of the actual output $x(t)$ of the system can be expressed as
\begin{equation} \label{eq1-5}
	X(s)=H(s)F(s)=H(s)F_m(s+\Omega),
\end{equation}
where $F_m(s)$ is the Laplace spectrum of $F_m(t)$.
Then, the Laplace spectrum $Z(s)$ of the mapped output $z(t)=T_{\Omega}^{-1}[x(t)]$ can be written as
\begin{equation} \label{eq1-6}
	Z(s)=X(s-\Omega)=H(s-\Omega)F_m(s) 
    = \dfrac{i \omega_n}{2 \sqrt{1-\zeta^2}} 
    \left[ \dfrac{1}{s - \Omega + \zeta \omega_n + i \sqrt{1-\zeta^2}\omega_n} - 
    \dfrac{1}{s - \Omega + \zeta \omega_n - i \sqrt{1-\zeta^2}\omega_n}
    \right] F_m(s).
\end{equation}
Note that when $\Omega  = \omega_n \zeta$ and $s\to i\omega_n\sqrt{1-\zeta^2}$, we have $|H(s-\Omega)|\to +\infty$. We refer to this $\Omega$ value as the optimal amplitude-decay factor for complex-frequency excitation, since the poles of the shifted transfer function $H(s-\Omega)$ lie on the imaginary axis, yielding an unbounded magnitude response. Accordingly, the mapped system's response $z(t)$ can be regarded as effectively undamped, which results in a  qualitatively distinct dynamic behavior.

\subsection{Information gain under white noise}
\label{Information gain under white noise}

% We consider a linear time-invariant damped resonator with a transfer function of 
% \begin{equation} \label{eq1-1}
% 	H(s) = \frac{\omega_n^2}{s^2+2\zeta\omega_ns+\omega_n^2},
% \end{equation}
% where $\omega_n$ is the natural frequency and $\zeta$ is the damping ratio. The complex-frequency operator $T_{\Omega}$ is defined as
% \begin{equation} \label{eq1-2}
% 	T_{\Omega}[f(t)]=e^{-\Omega t}f(t),
% \end{equation}
% where $f(t)$ is an arbitrary function of time $t$. Similarly, the inverse complex-frequency operator $T_{\Omega}^{-1}$ is defined as
% \begin{equation} \label{eq1-3}
% 	T_{\Omega}^{-1}[f(t)]=e^{\Omega t}f(t).
% \end{equation}

% Let the Laplace spectrum of the original excitation $f(t)$ be denoted as $f(s)$. Then, the Laplace spectrum of the modified excitation $F(t)=T_{\Omega}[f(t)]$ can be written as
% \begin{equation} \label{eq1-4}
% 	F(s)=f(s+\Omega).
% \end{equation}
% The Laplace spectrum $X(s)$ of the actual output $x(t)$ of the system can be expressed as
% \begin{equation} \label{eq1-5}
% 	X(s)=H(s)F(s)=H(s)f(s+\Omega).
% \end{equation}
% Then, the Laplace spectrum $Z(s)$ of the target output $z(t)=T_{\Omega}^{-1}[x(t)]$ can be written as
% \begin{equation} \label{eq1-6}
% 	Z(s)=X(s-\Omega)=H(s-\Omega)f(s).
% \end{equation}
% Therefore, the effective factor $Q_{\mathrm{eff}}$ of the system is calculated as
% \begin{equation} \label{eq1-7}
% 	Q_{\mathrm{eff}}(\Omega)=\frac{\omega_d}{2(\omega_n \zeta - \Omega)},
% \end{equation}
% where $\omega_d=\omega_n \sqrt{1-\zeta^2}$.

The mapping $T_{\Omega}^{-1}$ amplifies the system’s actual output, but it simultaneously amplifies the noise. The signal amplitude and thus the signal to noise ratio decreases exponentially in time for a constant ambient noise level. To quantify the role of noise on the frequency estimation accuracy, we determine the Fisher information of the system under our complex-frequency excitation framework. We assume the presence of Gaussian white noise, a common noise model with a flat power spectral density (PSD) that follows a Gaussian distribution in the time domain. 

Let $\omega_0$ be the frequency of $f(t)$, $x(t)$ be a measurement with the white noise $m(t)$ and 
$\sigma_m^2$ be the variance of noise at each measured time instant.
After the mapping $T_{\Omega}^{-1}$, the variance of the noise changes to
\begin{equation} \label{eq1-8}
	\mathrm{Var}\{e^{\Omega t_i}m_i \}=\sigma_m^2e^{2\Omega t_i}.
\end{equation}
Eq.~\eqref{eq1-8} can be rewritten in the form of matrix as 
$	\mathbf{\Sigma}_{\Omega}=\sigma_m^2\,\mathrm{diag}(e^{2\Omega t_i})$.
Let $\mathbf{z}_s$ be the vector of the ideal mapped output signal without the noise. The Fisher information $\mathcal{I}_{\Omega}$ is given by~\cite{steven1993fundamentals}
\begin{equation} \label{eq1-10}
	\mathcal{I}_{\Omega}=(\partial_{\omega_0} \mathbf{z}_s)^\top \mathbf{\Sigma}_\Omega^{-1}(\partial_{\omega_0} \mathbf{z}_s). 
\end{equation}
Noting that $e^{-\Omega t_i}z_i=x_{si}+ \sigma_m m_i$, Eq.~\eqref{eq1-10} can be rewritten as
\begin{equation} \label{eq1-12}
	\mathcal{I}_{\Omega}(\omega_0)=\frac{1}{\sigma_m^2}\left \| \partial_{\omega_0} \mathbf{x}_s  \right \| _2^2.
\end{equation}
Eq.~\eqref{eq1-12} indicates that the mapping $T_{\Omega}^{-1}$ does not introduce any additional Fisher information.

\begin{figure}[H]%fig1
	\makeatletter
	\renewcommand{\fnum@figure}{Fig. \thefigure.\@gobble}
	\makeatother
	\centering
	\includegraphics[scale=0.38]{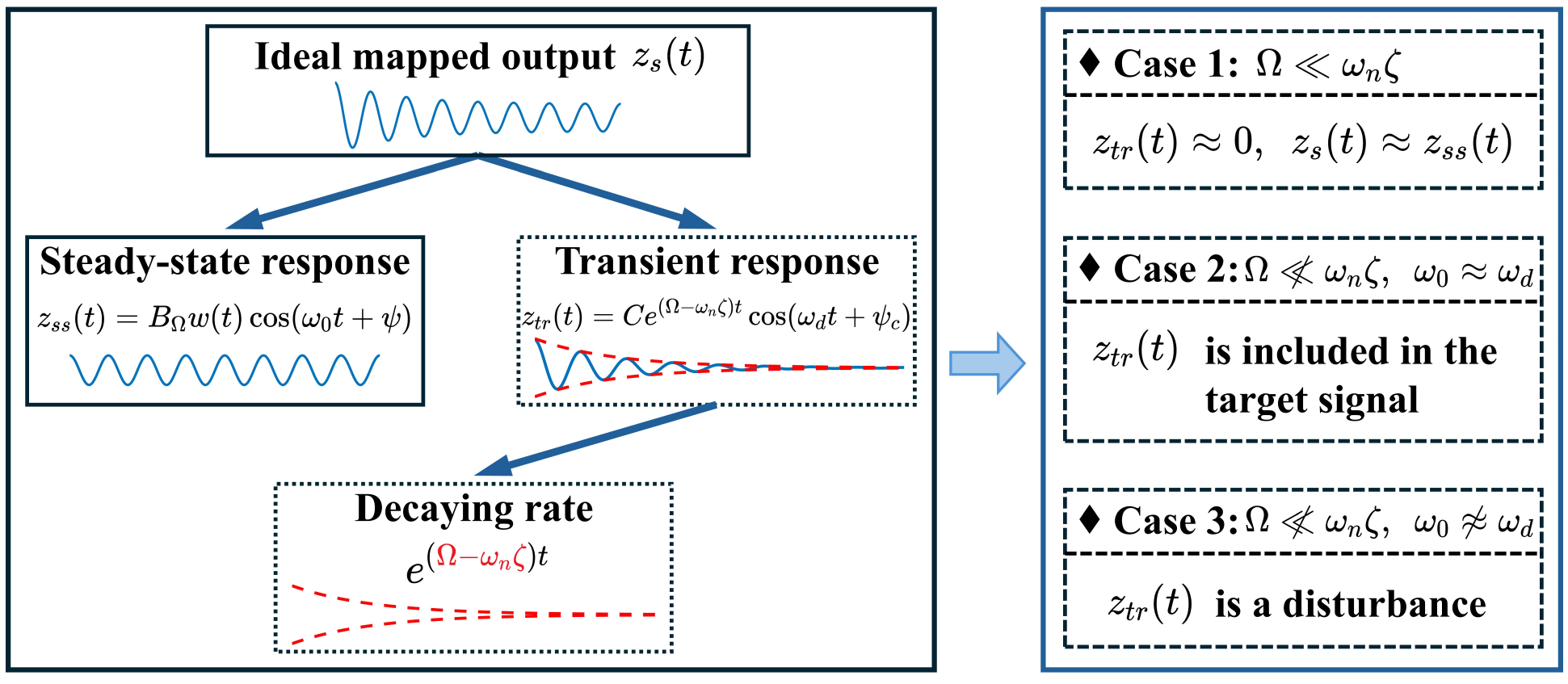}
	\caption{Three cases of transient-response effects on the Fisher information of the system under the complex-frequency excitation framework. The mapped output comprises a steady-state and a transient part. When $\Omega\ll\omega_n\zeta$, the transient response decays rapidly in the time domain and its influence on the Fisher information can be neglected. When $\Omega\not\ll\omega_n\zeta$, the relationship between the excitation frequency and the system’s damped natural frequency determines the specific impact of the transient response on the Fisher information.}
	\label{fig2-5}
\end{figure}

We discuss the approach to determine the Fisher information for a target frequency $\omega_0$. We will see that it depends on the relative value of $\omega_0$ and the damped natural frequency $\omega_d$, as well as on $\Omega$. Let us set $f(t)=Aw(t)\cos(\omega_0t+\phi)$, where $w(t)$ is a slowly varying window function. Then, the ideal mapped output $z_s(t)$ is expressed as
\begin{equation} \label{eq1-13}
	z_s(t)= z_{ss}(t) + z_{tr}(t) =B_{\Omega}w(t)\cos(\omega_0t+\psi)+Ce^{(\Omega-\omega_n\zeta)t}\cos(\omega_dt+\psi_c),
\end{equation}
where $C$ and $\psi_c$ are constants determined by initial conditions. 
A detailed derivation of $B_{\Omega}$ is provided in Appendix A. 
As can be seen from Eq.~\eqref{eq1-13}, $z_s(t)$ consists of a steady-state component $z_{ss}(t)$ and a transient component $z_{tr}(t)$. We always include $z_{ss}(t)$ in the target signal for frequency estimation. The role of $z_{tr}(t)$ in the frequency estimation process depends on the values of $\Omega$ and $\omega_0$, and is discussed in three separate cases. As shown in Fig.~\ref{fig2-5}, while $z_{ss}(t)$ has a constant amplitude, the amplitude of $z_{tr}(t)$ decays in time at a rate $e^{(\Omega-\omega_n\zeta)t}$. We first consider the simplest case, $\Omega\ll\omega_n\zeta$, for which $z_{tr}(t)$ decays rapidly in the time domain and can be neglected in frequency estimation, \textit{i.e.}, $z_s(t)\approx z_{ss}(t)$. When $\Omega\not\ll\omega_n\zeta$, the contribution of $z_{tr}(t)$ is no longer negligible. Since its oscillation frequency is the fixed value $\omega_d$ and is not necessarily equal to the steady-state frequency $\omega_0$, the impact of $z_{tr}(t)$ on frequency estimation requires further consideration. When $\omega_0\approx\omega_d$, $z_{tr}(t)$ can be regarded as part of the estimation target and yields a Fisher-information gain. In contrast, when $\omega_0 \not\approx \omega_d$, the transient component should be treated as an unknown disturbance rather than part of the target signal. It then acts as a nuisance component in the estimation problem and reduces the effective Fisher information. Therefore, these two cases must be analyzed separately.

Let us discuss the criterion to compare the Fisher information of complex-frequency and harmonic excitations. 
Since the amplitude of the complex-frequency excitation decays exponentially in time, its energy within the same time window is much smaller than that of the corresponding conventional excitation with $\Omega = 0$. We derive the Fisher information ratio under the assumption that the mapped excitation and the complex-frequency excitation have the same maximum amplitude, \textit{i.e.,} their amplitudes are equal at the initial time. Expressions for Fisher information ratio are also presented considering equal input energy. To this end, we introduce an energy-normalization factor $\Gamma(\Omega)$, which is determined by $w(t)$ and the excitation time $t_{final}$. $\Gamma$ is given by 
\begin{equation} \label{eq1-84}
	\Gamma(\Omega) \triangleq \frac{\int_0^{t_{final}}w^2(t)\cos ^2(\omega_0t+\psi)\mathrm{d}t}{\int_0^{t_{final}}e^{-2\Omega t}w^2(t)\cos ^2(\omega_0t+\psi)\mathrm{d}t}\approx\frac{\int_0^{t_{final}}w^2(t)\mathrm{d}t}{\int_0^{t_{final}}e^{-2\Omega t}w^2(t)\mathrm{d}t}.
\end{equation}

In the remainder of this section, we derive the Fisher-information gain under complex-frequency excitation for frequency estimation in three separate cases: when the transient response is negligible, when it is also included in the estimation target, and when it is treated as a linear disturbance.

\subsubsection{Case 1: Imaginary frequency far smaller than optimal ($\Omega\ll\omega_n\zeta$)}
\label{Imaginary frequency far from optimal}

In this case, the transient response of $z_s(t)$ decays rapidly as $e^{(\Omega-\omega_n\zeta) t}$ and its influence can be neglected. Hence, $\partial_{\omega_0} z_s(t)$ becomes
\begin{equation} \label{eq1-14}
	\partial_{\omega_0} z_s(t)=\partial_{\omega_0}[B_{\Omega}w(t)\cos(\omega_0t+\psi)]\approx-B_{\Omega}tw(t)\sin(\omega_0t+\psi).
\end{equation}
Here we assume that time of excitation and data measurement time are the same. We can rewrite Eq.~\eqref{eq1-10} from its discrete form into a continuous form as
\begin{equation} \label{eq1-15}
	\mathcal{I}_{\Omega}=\frac{1}{\sigma_m^2}\int_{0}^{t_{final}} e^{-2\Omega t}(\partial_{\omega_0} z_s(t))^2\mathrm{d}t,
\end{equation}
where $t_{final}$ denotes the excitation duration. Substituting Eq.~\eqref{eq1-14} into Eq.~\eqref{eq1-15} yields
\begin{equation} \label{eq1-16}
	\mathcal{I}_{\Omega}\approx\frac{B_{\Omega}^2}{2\sigma_m^2}J_{ss}(\Omega),
\end{equation}
where
\begin{equation*} \label{eq1-17}
	J_{ss}(\Omega)=\int_{0}^{t_{final}} e^{-2\Omega t}w(t)^2t^2\mathrm{d}t.
\end{equation*}
Therefore, the Fisher-information ratio can be expressed as
\begin{equation} \label{eq1-18}
	R_{\mathcal{I}}(\Omega)=\frac{\mathcal{I}_{\Omega}}{\mathcal{I}_{0}}\approx\left ( \frac{B_{\Omega}}{B_0} \right ) ^2\frac{J_{ss}(\Omega)}{J_{ss}(0)}=\left ( \frac{|H(s-\Omega)|}{|H(s)|} \right ) ^2\frac{J_{ss}(\Omega)}{J_{ss}(0)}.
\end{equation}
After energy normalization, the equivalent Fisher information ratio $R_{\mathcal{I}}^{(eq)}(\Omega)$ can be obtained as
\begin{equation} \label{eq1-85}
	R_{\mathcal{I}}^{(eq)}=\Gamma(\Omega)R_{\mathcal{I}}(\Omega)\approx\Gamma(\Omega)\left ( \frac{B_{\Omega}}{B_0} \right ) ^2\frac{J_{ss}(\Omega)}{J_{ss}(0)}.
\end{equation}

\begin{figure}[H]%fig1
	\makeatletter
	\renewcommand{\fnum@figure}{Fig. \thefigure.\@gobble}
	\makeatother                    
	\centering
	\includegraphics[scale=0.65]{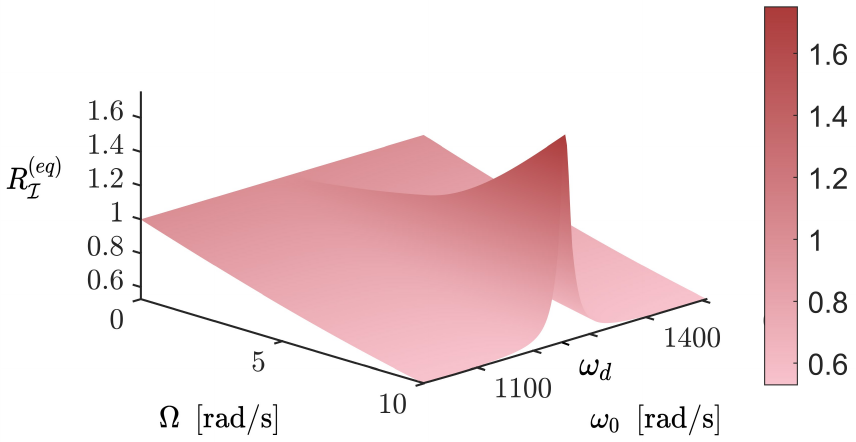}
	\caption{Dependence of the equivalent Fisher information ratio $R_{\mathcal{I}}^{(eq)}$ on the amplitude-decay factor $\Omega$ and the excitation frequency $\omega_0$ when $\Omega\ll\omega_n\zeta$. When $\omega_0$ is close to $\omega_d$, $R_{\mathcal{I}}^{(eq)}$ increases monotonically with $\Omega$; when $\omega_0$  is far from $\omega_d$ , $R_{\mathcal{I}}^{(eq)}$ decreases monotonically with $\Omega$.}	
	\label{fig2-2}
\end{figure}

We illustrate the dependence of $R_{\mathcal{I}}^{(eq)}$ on excitation parameters with an example, setting $\omega_n=1257$ [rad/s], $\zeta=0.02$, and $t_{final}=0.2$ s. Figure~\ref{fig2-2} displays how $R_{\mathcal{I}}^{(eq)}$ varies with the real $\omega_0$ and imaginary $\Omega$ components of the complex frequency. When $\omega_0$ is far from $\omega_d$, $R_{\mathcal{I}}^{(eq)}$ decreases gradually as $\Omega$ increases. By contrast, when $\omega_0$ approaches $\omega_d$, $R_{\mathcal{I}}^{(eq)}$ increases with increasing $\Omega$.

\subsubsection{Case 2: Target frequency close to damped natural frequency ($\omega_0\approx\omega_d$)} 
When $\Omega$ is no longer much smaller than $\omega_n \zeta$, the transient response decays sufficiently slowly that its contribution to the Fisher information can no longer be neglected. Although the transient-response frequency $\omega_d$ is independent of the excitation frequency $\omega_0$, two distinct situations arise depending on their separation. When $\omega_0 \approx \omega_d$, the transient and steady-state components become nearly frequency-matched and are difficult to distinguish; the transient can then be treated as an additional coherent component of the target signal. In this case, it can increase the Fisher information, with the amount of increase depending on the phase alignment between the two components. 

Here, both the steady-state and transient components of $z_s(t)$ contain Fisher information about $\omega_0$. Setting $\omega_d \approx \omega_0$, Eq.~\eqref{eq1-13} becomes
\begin{equation*} \label{eq1-42}
	z_s(t)\approx B_{\Omega}w(t)\cos(\omega_0t+\psi)+Ce^{(\Omega-\omega_n\zeta )t}\cos(\omega_0t+\psi_c).
\end{equation*}
When $\Omega=\omega_n\zeta$, $H(s - \Omega)$ becomes an undamped transfer function. Consequently, $B_{\Omega}$ is no longer a constant equal to $A|H(j\omega_0 - \Omega)|$; instead, $B_{\Omega}$ becomes a function of time. 
At this point, $\partial_{\omega_0} z_s(t)$ is calculated as
\begin{equation*} \label{eq1-43}
	\partial_{\omega_0} z_s(t)=-t\Big[B_{\Omega}w(t)\sin(\omega_0t+\psi)+Ce^{(\Omega-\omega_n\zeta )t}\sin(\omega_0t+\psi_c)\Big].
\end{equation*}
Based on Eq.~\eqref{eq1-15}, we have
\begin{equation} \label{eq1-44}
\begin{split}
	\mathcal{I}_{\Omega} &= \frac{1}{\sigma_m^2}\int_0^{t_{final}}e^{-2\Omega t}t^2 \Big[ B_{\Omega}^2w(t)^2\sin^2(\omega_0t + \psi) + C^2e^{2(\Omega-\omega_n\zeta )t}\sin^2(\omega_0 t + \psi_c) \\
	&\quad + 2B_{\Omega}Ce^{(\Omega-\omega_n\zeta )t}w(t)\sin(\omega_0t + \psi)\sin(\omega_0 t + \psi_c) \Big].
\end{split}
\end{equation}
Noting that 
$2 \sin(\omega_0 t + \psi)\sin(\omega_0 t + \psi_c)=[ \cos(\psi-\psi_c)-\cos(2\omega_0t+\psi+\psi_c) ], 
$ and setting 
\begin{equation*} \label{eq1-46}
	\rho\triangleq\cos(\psi-\psi_c), 
\end{equation*}
Eq.~\eqref{eq1-44} can be rewritten as
\begin{equation*} \label{eq1-47}
	\mathcal{I}_{\Omega} = \frac{1}{2\sigma_m^2} \Big[B_{\Omega}^2J_{ss}(\Omega)+C^2J_{tr}+2\rho B_{\Omega}CJ_{mix}(\Omega) \Big],
\end{equation*}
% \begin{equation} \label{eq1-48}
% 	J_{tr}=\int_0^{t_{final}}t^2e^{-2\omega_n\zeta\, t}\mathrm{d}t,
% \end{equation}
% \begin{equation} \label{eq1-49}
% 	J_{mix}(\Omega)=\int_0^{t_{final}}t^2e^{-(\Omega+\omega_n\zeta) t}w(t)\mathrm{d}t.
% \end{equation}
\begin{equation*} \label{eq1-48}
	J_{tr}=\int_0^{t_{final}}t^2e^{-2\omega_n\zeta\, t}\mathrm{d}t,
\qquad	J_{mix}(\Omega)=\int_0^{t_{final}}t^2e^{-(\Omega+\omega_n\zeta) t}w(t)\mathrm{d}t.
\end{equation*}
The Fisher-information ratio can then be expressed as
\begin{equation} \label{eq1-50}
	R_{\mathcal{I}}(\Omega)=\frac{\mathcal{I}_{\Omega}}{\mathcal{I}_{0}}\approx \frac{B_{\Omega}^2J_{ss}(\Omega)+C^2J_{tr}+2\rho B_{\Omega}CJ_{mix}(\Omega)}{{B_0}^2J_{ss}(0)}.
\end{equation}
Eq.~\eqref{eq1-50} indicates that the Fisher information ratio under complex-frequency excitation depends not only on the total excitation duration $t_{final}$ and the complex-frequency amplitude-decay factor $\Omega$, but also on the system’s initial state. When the transient response amplitude becomes larger and the phase difference between the transient and steady-state components becomes smaller, the Fisher information increases. The parameters $C$ and $\psi_c$ are determined by the initial conditions $x(t=0)$, $\dot{x}(t=0)$ and are assumed to be noise-free. The physical intuition here is that increasing the amplitude of the transient response enhances the energy of the target signal, while the energy of the Gaussian noise remains unchanged, leading to an increase in the Fisher information.
Finally, the equivalent Fisher information ratio $R_{\mathcal{I}}^{(eq)}(\Omega)$ is then
\begin{equation} \label{eq1-86}
	R_{\mathcal{I}}^{(eq)}\approx\Gamma(\Omega)\frac{B_{\Omega}^2J_{ss}(\Omega)+C^2J_{tr}+2\rho B_{\Omega}CJ_{mix}(\Omega)}{{B_0}^2J_{ss}(0)}.
\end{equation}

\begin{figure}[H]%fig1
	\makeatletter
	\renewcommand{\fnum@figure}{Fig. \thefigure.\@gobble}
	\makeatother
	\centering
	\includegraphics[scale=0.7]{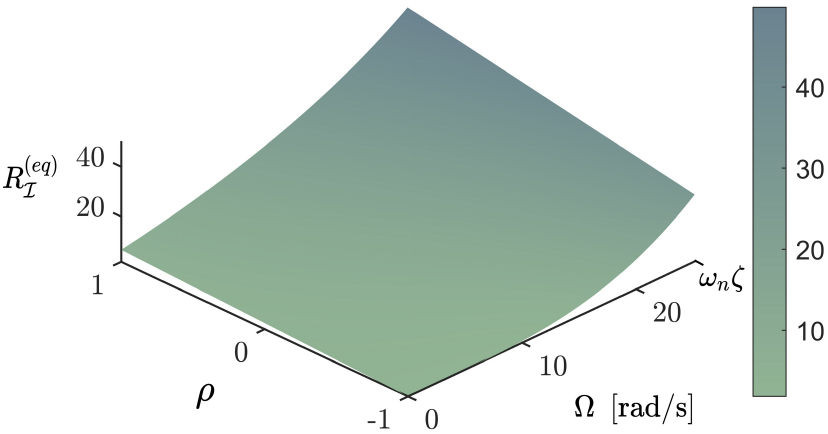}
	\caption{Dependence of the equivalent Fisher information ratio $R_{\mathcal{I}}^{(eq)}$ on the amplitude-decay factor $\Omega$ and $\rho$ when $\omega_0\approx\omega_d$. $R_{\mathcal{I}}^{(eq)}$ increases monotonically with $\Omega$; moreover, as $\rho$ approaches unity, \textit{i.e.,} as the phase difference between the steady state and transient components decreases, $R_{\mathcal{I}}^{(eq)}$ becomes larger.}
	\label{fig2-3}
\end{figure}

 The values of the parameters $\omega_n$, $\zeta$, and $t_{final}$ are kept the same as those in Sec.~\ref{Imaginary frequency far from optimal}. Based on Eq.~\eqref{eq1-86}, the dependence of $R_{\mathcal{I}}^{(eq)}$ on $\Omega$ and $\rho$ is illustrated in Fig. \ref{fig2-3}. It indicates that $R_{\mathcal{I}}^{(eq)}$ increases as the phase alignment between the steady-state and transient components becomes closer, \textit{i.e.},  as $\rho$ approaches unity. In addition, $R_{\mathcal{I}}^{(eq)}$ increases monotonically with $\Omega$, since in this regime, $R_{\mathcal{I}}^{(eq)}$ is dominated by the transient term $J_{\mathrm{tr}}$. This term is independent of $\Omega$ but is amplified by the energy-normalization factor $\Gamma(\Omega)$. Since $\Gamma(\Omega)$ increases monotonically with $\Omega$, $R_{\mathcal{I}}^{(eq)}$ correspondingly grows with $\Omega$.

% Next, we extend the discussion to the case of $\omega_0 = \omega_d$ and $\Omega \not \approx \omega_n \zeta$. In this situation, the expression of the information gain ratio is identical to that in Eq. (\ref{eq1-42}) and can be represented in terms of the transfer function as
% \begin{equation} \label{eq1-51}
%     \begin{split}
% 	R_{\mathcal{I}}(\Omega)=\frac{\mathcal{I}_{\Omega}}{\mathcal{I}_{0}} \approx \frac{B_{\Omega}^2J_{ss}(\Omega)+C^2J_{tr}(\Omega)+2\rho B_{\Omega}CJ_{mix}(\Omega)}{{B_0}^2J_{ss}(0)}\\
%     =\frac{|H(j\omega_0-\Omega)|^2J_{ss}(\Omega)+C^2J_{tr}(\Omega)+2\rho B_{\Omega}CJ_{mix}(\Omega)}{|H(j\omega_0)|^2J_{ss}(0)}
%     \end{split}.
% \end{equation}

\subsubsection{Case 3: Target frequency far from damped natural frequency ($\omega_0\not \approx\omega_d$)}
Finally, we analyze the case in which the transient-response frequency is no longer close to the steady-state response frequency, so that the transient response is treated as an additional noise source. The Fisher information undergoes a substantial change because the transient response is then treated as an unknown linear disturbance, which adversely affects the estimation of $\omega_0$. The mapped output $z(t)$ can be rewritten as
\begin{equation} \label{eq1-51}
	z(t)=z_{ss}(t)+z_{tr}(t)+n(t),
\end{equation}
where $z_{ss}(t)$ is the steady-state response, $z_{tr}(t)$ denotes the transient response, and $n(t)$ is the noise obtained by applying $T_m = T_{\Omega}^{-1}$ to the original Gaussian white noise. The terms $z_{ss}(t)$ and $z_{tr}(t)$ can be expressed as
\begin{equation} \label{eq1-52}
	z_{ss}(t)=B_{\Omega}w(t)\cos(\omega_0t+\psi),
\end{equation}
\begin{equation} \label{eq1-53}
	z_{tr}(t)=\alpha e^{(\Omega-\omega_n\zeta)t}\cos(\omega_dt)+\beta e^{(\Omega-\omega_n\zeta)t}\sin(\omega_dt),
\end{equation}
where $\alpha=C\cos\psi_c$ and $\beta=-C\sin\psi_c$. At this stage, the values of $\alpha$ and $\beta$ influence the estimation of $\omega_0$. 
The detailed derivation is provided in Appendix B. 
After derivation, the Fisher information can be expressed as
\begin{equation*} \label{eq1-82}
	\mathcal{I}_{\Omega}\approx\frac{B_{\Omega}^2}{2\sigma_m^2}\left[ J_{ss}(\Omega)-\frac{4L_{mix}^2(\Omega)}{L_{tr}}\left( r_1^2(\Omega)+r_2^2(\Omega)\right) \right],
\end{equation*}
where
\begin{equation*} \label{eq2-2-1}
	L_{tr}=\int_0^{t_{final}}e^{-2\omega_n\zeta t}\mathrm{d}t,
\end{equation*}
\begin{equation*} \label{eq2-2-2}
	L_{mix}(\Omega)= \int_0^{t_{final}}te^{-(\Omega+\omega_n\zeta) t}w(t)\mathrm{d}t,
\end{equation*}
\begin{equation*} \label{eq2-2-3}
	r_1(\Omega)=\dfrac{1}{L_{mix}(\Omega)}\int_0^{t_{final}}te^{-(\Omega+\omega_n\zeta) t}w(t)\cos(\omega_dt)\sin(\omega_0t+\psi)\mathrm{d}t,
\end{equation*}
\begin{equation*} \label{eq2-2-4}
	r_2(\Omega)=\dfrac{1}{L_{mix}(\Omega)}\int_0^{t_{final}}te^{-(\Omega+\omega_n\zeta) t}w(t)\sin(\omega_dt)\sin(\omega_0t+\psi)\mathrm{d}t.
\end{equation*}

At this point, the Fisher information ratio can be expressed as
\begin{equation} \label{eq1-83}
	R_{\mathcal{I}}(\Omega)=\frac{\mathcal{I}_{\Omega}}{\mathcal{I}_{0}}\approx\frac{B_{\Omega}^2J_{ss}(\Omega)-4B_{\Omega}^2\left( L_{mix}^2(\Omega)/L_{tr}\right)\left( r_1^2(\Omega)+r_2^2(\Omega)\right)}{B_0^2J_{ss}(0)}.
\end{equation}
Eq.~\eqref{eq1-83} shows that when $\omega_0 \not \approx \omega_d$, the transient response acts as an unknown linear disturbance and leads to a reduction in the Fisher information. Moreover, the amount of reduction is determined solely by the normalized correlation between the transient subspace and the derivative of the steady-state response, and is independent of the transient amplitude. 
Then, the equivalent Fisher information ratio $R_{\mathcal{I}}^{(eq)}(\Omega)$ can be calculated as
\begin{equation} \label{eq1-87}
	R_{\mathcal{I}}^{(eq)}\approx\Gamma(\Omega)\frac{B_{\Omega}^2J_{ss}(\Omega)-4B_{\Omega}^2\left( L_{mix}^2(\Omega)/L_{tr}\right)\left( r_1^2(\Omega)+r_2^2(\Omega)\right)}{B_0^2J_{ss}(0)}.
\end{equation}

\begin{figure}[H]%fig1
	\makeatletter
	\renewcommand{\fnum@figure}{Fig. \thefigure.\@gobble}
	\makeatother
	\centering
	\includegraphics[scale=0.7]{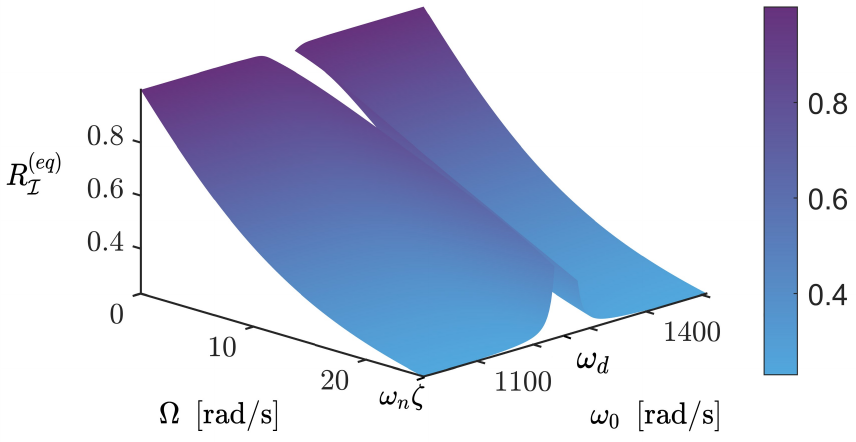}
	\caption{Dependence of the equivalent Fisher information ratio $R_{\mathcal{I}}^{(eq)}$ on the amplitude-decay factor $\Omega$ and the excitation frequency $\omega_0$ when $\omega_0\not \approx\omega_d$. The value of $R_{\mathcal{I}}^{(eq)}$ decreases as $\Omega$ increases, and, since the transient component is treated as a disturbance in this case, it decreases more rapidly than in Fig. \ref{fig2-2}. For a fixed $\Omega$, $R_{\mathcal{I}}^{(eq)}$ increases gradually as $\omega_0$ approaches $\omega_d$.
}	
	\label{fig2-4}
\end{figure}

Similarly, the values of the parameters $\omega_n$, $\zeta$, and $t_{final}$ are kept the same as those in Sec.~\ref{Imaginary frequency far from optimal}. In this case, the dependence of $R_{\mathcal{I}}^{(eq)}$ on $\Omega$ and $\omega_0$ is displayed in Fig. \ref{fig2-4}. We see from the figure that  $R_{\mathcal{I}}^{(eq)}$ decreases with increasing $\Omega$. This behavior arises because, when $\omega_0 \not \approx \omega_d$, the increase in $B_{\Omega} $ is insufficient to compensate for the reduction in $\Gamma(\Omega) J_{ss}(\Omega)$; moreover, the projection of the transient term is treated as an interference component, which further decreases $R_{\mathcal{I}}^{(eq)}$. However, for a fixed $\Omega$, as $\omega_0$ approaches $\omega_d$, the increase in $B_{\Omega}$ becomes more pronounced, and $R_{\mathcal{I}}^{(eq)}$ correspondingly increases.

% Eq. (\ref{eq1-82}) provides a general conclusion for the case $\omega_0 \not \approx \omega_d$. When the excitation is narrowband and its frequency is close to the resonance frequency, Eq. (\ref{eq1-82}) can be further simplified.

\subsection{Information gain under colored noise}\label{sec:colored_theory}
While Gaussian colored noise exhibits a Gaussian amplitude distribution in the time domain, in contrast to Gaussian white noise, its power spectral density (PSD) is not flat. For example, the PSDs of pink noise and red noise are inversely proportional to frequency and to the square of frequency, respectively. In practical applications, some noise sources arise in the form of Gaussian colored noise \cite{dutta1981low}. In micro- and nanoscale sensing, for example, pink noise commonly coexists with white noise due to material defects and impurities in the sensor structure \cite{burnett2014evidence,sansa2016frequency}. The mapping $T_{\Omega}^{-1}$ likewise amplifies colored noise, thereby affecting the accuracy of the frequency-estimation results. Therefore, we elucidate how the noise power spectral distribution influences the Fisher information of the system under complex-frequency excitation in this section.

When the variation of the colored-noise PSD $S_m(\omega)$ over the target narrowband $\Xi$ is negligible, the Fisher information ratio and the equivalent Fisher information ratio take the same forms as those in Sec. \ref{Information gain under white noise}. When $S_m(\omega)$ varies significantly over $\Xi$ and can no longer be approximated as constant, the Fisher information expression can be rewritten as
\begin{equation*} \label{eq2-3-1}
	\mathcal{I}_{\Omega}\approx\frac{1}{2\pi}\int_{\Xi}\frac{|\partial _{\omega_0}X_{\Omega} (j\omega) |^2}{S_m(\omega)}\mathrm{d}\omega.
\end{equation*}
Here $X_{\Omega} (j\omega)$ is the spectrum of the ideal noise-free actual output. In this case, relative to the white-noise scenario, the Fisher information under colored noise is weighted in the frequency domain by $1/S_m(\omega)$. Because the explicit form of $S_m(\omega)$ is generally unknown, a closed-form expression for the Fisher information ratio analogous to that in Sec. \ref{Information gain under white noise} cannot be provided. Nevertheless, the conclusions in Sec. \ref{Information gain under white noise} regarding the influence of the system’s initial state on the information ratio remain valid, because the Fisher-information variation induced by the initial state is independent of the noise power spectral distribution. Detailed derivations and explicit expressions  of the Fisher information ratio in this section are provided in Appendix C.

\section{Numerical simulations}
\label{Numerical simulations}
% Under both white and colored noise conditions, a second-order linear time-invariant system is numerically constructed, and Monte Carlo simulations are performed to evaluate the variance of frequency estimation for the target output under complex-frequency excitation. The obtained variance is then compared with the theoretical predictions to validate the correctness of the proposed theory.

% In this simulation, the target excitation is specified as
% \begin{equation} \label{eq1-84}
% 	f(t)=Aw(t)\cos(\omega_0t+\varphi),
% \end{equation}
% where $w(t)$ is a slowly varying Tukey window in time. By applying the complex-frequency excitation operator to $F_{\mathrm{ori}}(t)$, we obtain
% \begin{equation} \label{eq1-85}
% 	F(t)=T_{\Omega}\left [ f(t) \right ]=e^{-\Omega t}\cdot f(t).
% \end{equation}
% Then, by performing discrete convolution with the system response $h(t)$, the noise-free actual output $x_s(t;\omega_0,\Omega)$ is obtained as
% \begin{equation} \label{eq1-86}
% 	x_s(t;\omega_0,\Omega)=F(t;\omega_0,\Omega)*h(t).
% \end{equation}
To verify the theoretical predictions of Fisher information ratio derived in Sec.~\ref{Theory}, we perform Monte Carlo simulations. We first generate the discrete actual response signal and add Gaussian white noise. The noisy actual response is then subjected to the mapping $T_m$ to obtain the mapped response. The frequency is estimated by projecting this mapped response onto the signal subspace associated with the corresponding parametric response model and maximizing the resulting weighted projection energy over the trial frequency. The sample variance of the estimates is then compared with the theoretical predictions.

\subsection{Simulation Procedure}

Let the sampling interval be $\Delta t$, and the total simulation time be $t_{final}$. The discrete time sequence can thus be expressed as
\begin{equation} \label{eq2-1}
	t_n=n\Delta t,\ \ \ \ n=0,\ 1,...,\ N,
\quad \text{with  }	N = t_{final} / \Delta t.
\end{equation}
All signals are measured at times $\{t_n\}_{n=0}^{N}$. 
We set the following windowed narrowband excitation as the original input:
\begin{equation} \label{eq2-3}
	 f(t) = A\, w(t)\cos(\omega_0 t + \phi),
\end{equation}
where $A$ is the amplitude of the original excitation, $\omega_0$ is the frequency close to the natural frequency of the system, $\phi$ is the initial phase of the original excitation, and $w(t)$ is the window function. Its discrete form can be written as 
\begin{equation} \label{eq2-4}
	 f[n] = A\, w[n]\cos(\omega_0 t_n + \phi),
\end{equation}
In the theoretical derivations in Sec.~\ref{Theory}, $w[n]$ is only required to be a slowly varying function; therefore, for convenience, a Tukey window is employed in the present simulation. It is given by \cite{harris1978use}:
\begin{equation} \label{eq2-5}
	 w[n] = \begin{dcases}
\frac{1}{2} - \frac{1}{2}\cos \left( 2\pi \frac{n}{\alpha N} \right) & 0 \le n < \frac{\alpha N}{2}, \\
1 & \frac{\alpha N}{2} \le n < N - \frac{\alpha N}{2} \\
\frac{1}{2} - \frac{1}{2}\cos \left( 2\pi \frac{N-n}{\alpha N} \right) & N - \frac{\alpha N}{2} \le n < N.
\end{dcases},
\end{equation}
Here $\alpha$ is a parameter that controls the flatness of the window function.
The specific choice of the window function $w(t)$ does not affect this verification, because the theoretical results already account for the influence of the window on the Fisher information. It suffices that the window satisfies the basic assumption of being slowly varying in time domain.

The complex-frequency excitation $F(t)$ is obtained by applying the mapping $T_{\Omega}$ to the original excitation $f(t)$:
\begin{equation} \label{eq2-6}
    F(t)=T_{\Omega}[f(t)]=A\ e^{-\Omega t}\ w(t)\cos(\omega_0t+\phi).
\end{equation}
The continuous-time impulse response $h(t)$ of this system is
\begin{equation} \label{eq2-9}
    h(t)=\frac{1}{\sqrt{1-\zeta^2}}\ e^{-\zeta\omega_n t}\sin(\omega_d t).
\end{equation}
%where $\omega_d=\omega_n\sqrt{1-\zeta^2}$. 
and the corresponding actual output $x(t)$ is given by~\cite{inman1994engineering} 
\begin{equation*} \label{eq2-10}
    x(t) = \int_{0}^t h(\tau)  F(t-\tau) d\tau. 
\end{equation*}
In this simulation, a noise $m[n]\sim \mathcal N(0,\sigma_m^2)$ is added to the actual output $x[n]$. Then, the observed actual output $\tilde{x}[n]$ is $\tilde{x}[n]=x[n]+m[n]$. The discrete form $\tilde{z}[n]$ of the mapped output is then given by 
\begin{equation} \label{eq2-14}
    \tilde{z}[n]=T_{m}(\tilde{x}[n])=e^{\Omega t_n}( x[n]+m[n]) .
\end{equation}

To determine the frequency, a least squares-based estimator is used here. The idea is to fit the observed mapped output sequence $\tilde{z}[n]$ with a narrowband harmonic signal with a known envelope. Therefore, for each trial frequency $\omega$, the basis function sequences are defined as
\begin{equation*} 
    s_1[n;\omega] = a[n] \cos(\omega t_n), \qquad
    s_2[n;\omega] = a[n] \sin(\omega t_n),
\end{equation*}
where $a[n]$ is the equivalent envelope sequence, given below. By stacking $s_1$ and $s_2$, the basis matrix $\mathbf{S}=[\mathbf{s}_1,\,\mathbf{s}_2]$ is constructed. 
% Since the signal model used to compute the Fisher information differs for each information-gain-ratio result, the specific form of $a[n]$ is not fixed. 
For $R_{\mathcal{I}}$ $(\Omega\ll\omega_n\zeta)$ and $(\omega_0\not \approx\omega_d)$, since the transient component does not contain any Fisher information, $a[n]$ can be expressed as 
\begin{equation} \label{eq2-18}
    a[n] = B_{\Omega}w[n].
\end{equation}
However, for $R_{\mathcal{I}}$ $(\omega_0\approx\omega_d)$, since the transient component is also used to calculate the Fisher information, $\mathbf{S}$ needs to be expanded as
\begin{equation} \label{eq2-19}
    \mathbf{S}=\begin{bmatrix}
               \mathbf{s}_1 & \mathbf{s}_2 & \mathbf{s}_3 & \mathbf{s}_4
                \end{bmatrix},
\end{equation}
\begin{equation} \label{eq2-19-1}
    s_3[n;\omega] = Ce^{(\Omega-\omega_n\zeta )t_n} \cos(\omega t_n),
\qquad
    s_4[n;\omega] = Ce^{(\Omega-\omega_n\zeta )t_n} \sin(\omega t_n).
\end{equation}
The projection energy $E(\omega)$ is then obtained via weighted least squares \cite{scharf1991statistical}. It is given by 
\begin{equation} \label{eq2-20}
    E(\omega)=\tilde{\mathbf{z}}^{\top}\mathbf{W}\mathbf{S}(\omega)\left ( \mathbf{S}^{\top}(\omega)\mathbf{W} \mathbf{S}(\omega) \right )^{-1}\mathbf{S}^{\top}(\omega)\mathbf{W}\tilde{\mathbf{z}} .
\end{equation}
%\begin{equation} \label{eq2-20-1}
%    \mathbf{W}=\mathrm{diag}\left (e^{-\Omega t_n} \right ) .
%\end{equation}
Here $\mathbf{W}=\mathrm{diag}\left (e^{-\Omega t_n} \right )$ is a whitening matrix, whose role is to transform the exponentially increasing noise in $\tilde{\mathbf{z}}$ into Gaussian white noise with constant mean in time, thereby facilitating frequency estimation. Since maximizing the likelihood under Gaussian noise is equivalent to maximizing the weighted projection energy, the value of $\omega$ that maximizes $E(\omega) $ is taken as the frequency estimate $\hat{\omega}_0$. Accordingly, $\hat{\omega}^{(q)}_0$ can be expressed as
\begin{equation} \label{eq2-21}
    \hat{\omega}_0=\arg \underset{\omega\in \mathcal{S}}{\max}E(\omega),
\end{equation}
where $\mathcal{S} =  \left[\omega_{\min},\,\omega_{\max} \right]$ is a narrow frequency interval. Note that this interval has a distinct physical meaning from the frequency interval in Sec.~\ref{sec:colored_theory}. Here, it  refers to a small interval within which the frequency corresponding to the maximum projection energy may appear. By contrast, in Sec.~\ref{sec:colored_theory}, it refers to a narrow frequency band of interest that we assume a priori, within which we discuss variations in the noise power spectral density, centered around $\omega_0$. 

To verify the Fisher information ratio expressions derived in Sec.~\ref{Theory}, we compute the ratio of variances obtained from numerical simulations. 
For each $\Omega$, the above simulation procedure is repeated $Q=1000$ times to compute the sample variance $\mathrm{Var}_{\Omega}\left [\hat{\omega}_0\right ]$, and the variance ratio is then constructed as
\begin{equation} \label{eq2-22}
    R_{\mathcal{I}}(\Omega)=\frac{\mathrm{Var}_{0}\left [\hat{\omega}_0\right ]}{\mathrm{Var}_{\Omega}\left [\hat{\omega}_0\right ]}.
\end{equation}
The variance ratio $ R_{\mathcal{I}}$ obtained from numerical simulations is compared with theoretical $R_{\mathcal{I}}$ in Sec. \ref{Information gain under white noise} to verify those derived expressions.

For the verification of $R_{\mathcal{I}}$ $(\omega_0\not \approx\omega_d)$, because the transient response is treated as an additional linear disturbance, the computation of the projection energy differs from that in Eq. \eqref{eq2-20}. Based on the basis matrix $\mathbf{K}$ in Eq. \eqref{eq1-59}, its orthogonal-complement projection matrix can be calculated as
\begin{equation*} \label{eq2-23}
    P_{\mathbf{K}}^{\bot}=\mathbf{I}-\mathbf{K}\left( \mathbf{K}^{\top}\mathbf{W}\mathbf{K} \right )^{-1}\mathbf{K}^{\top}\mathbf{W}.
\end{equation*}
Then the projection energy can be rewritten as
\begin{equation} \label{eq2-24}
    E(\omega)=\left (P_{\mathbf{K}}^{\bot}\tilde{\mathbf{z}}^{(q)} \right )^{\top}P_{\mathbf{K}}^{\bot}\mathbf{S}(\omega)\left ( \left(P_{\mathbf{K}}^{\bot}\mathbf{S}(\omega)\right)^{\top}P_{\mathbf{K}}^{\bot} \mathbf{S}(\omega) \right )^{-1}\left(P_{\mathbf{K}}^{\bot}\mathbf{S}(\omega)\right)^{\top}P_{\mathbf{K}}^{\bot}\tilde{\mathbf{z}}^{(q)}.
\end{equation}

\subsection{Simulation parameters and numerical results}

\begin{figure}[H]%fig1
	\makeatletter
	\renewcommand{\fnum@figure}{Fig. \thefigure.\@gobble}
	\makeatother
	\centering
	\includegraphics[scale=0.55]{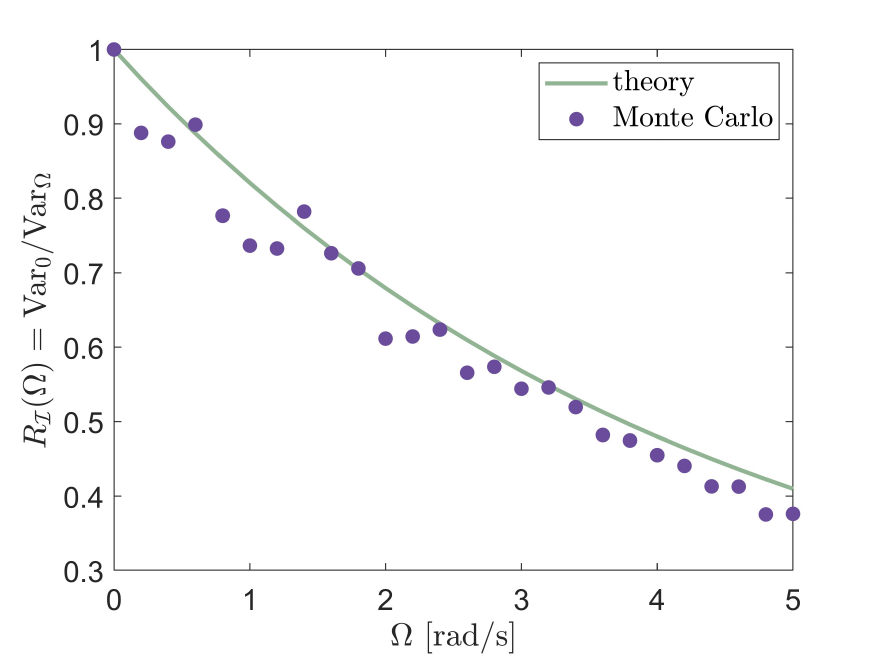}
	\caption{$R_{\mathcal{I}}$ obtained from simulations and theory for the case $\Omega\ll\omega_n\zeta$. The simulation results agree closely with the theoretical predictions. With all other parameters held constant, increasing $\Omega$ leads to a more pronounced amplification of noise, thereby resulting in a decrease of the Fisher information ratio.}
	\label{fig3-1}
\end{figure}

The parameters used in the simulations are listed in Table 1. First, we consider the case  $(\Omega\ll\omega_n\zeta)$. In the simulation for $R_{\mathcal{I}}$, we set $\omega_0=\omega_n\sqrt{1-2\zeta^2}=1255.7\ \mathrm{rad/s}$, and vary $\Omega$ from 0 to 5 rad/s with a step size of 0.2 rad/s. Figure~\ref{fig3-1} displays the results. When $\Omega$ is significantly smaller than $\omega_n \zeta$, the value of $R_{\mathcal{I}}$ obtained from Monte Carlo simulations agrees very well with the theoretical prediction of Eq.~\eqref{eq1-18}, thereby verifying the correctness of the derived expression.
\begin{table}[ht]
\renewcommand\arraystretch{1}
\centering
\caption{System and excitation parameters in the simulation.}
\begin{tabular}{llll}
\hline
 Symbol & Value & Unit   \\ \hline
 $\Delta t$    & $5\times10^{-6}$  & s     \\
 $\omega_n$    & 1257    & rad/s     \\
 $\zeta$    & 0.02 & -    \\
 $t_{final}$    & 0.2   & s \\
 $\alpha$  & 0.5 & - \\
 $A$ & 10 & -\\
 $\sigma_m$ & 1 & -\\ \hline
\end{tabular}
\end{table}

\begin{figure}[H]%fig3
	\makeatletter
	\renewcommand{\fnum@figure}{\textbf{Fig.} \thefigure.\@gobble}
	\makeatother
	\subfigure[]
	{
		\begin{minipage}[t]{0.5\linewidth}
			%\flushright
			\centering
			\includegraphics[scale=0.55]{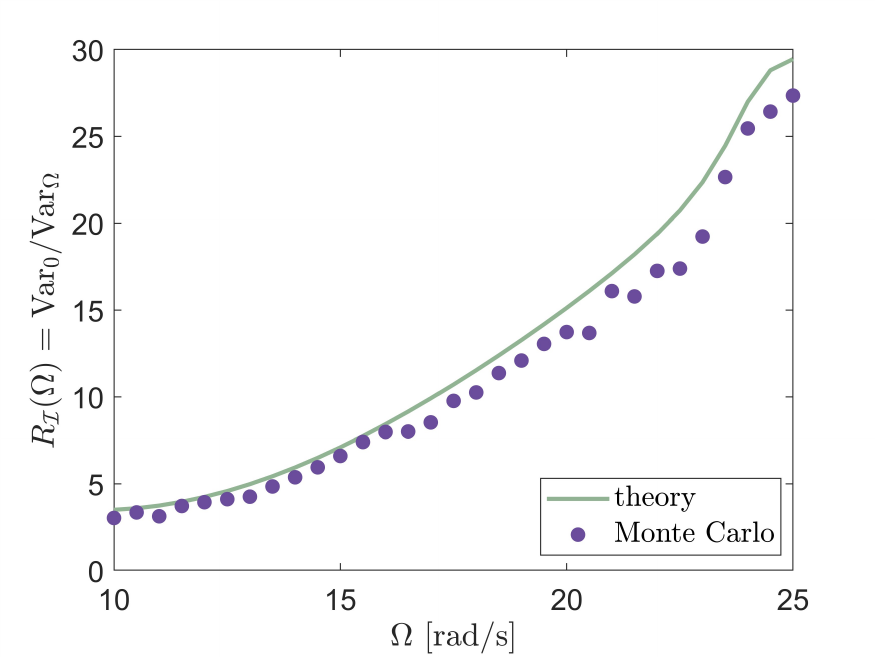}
			\label{fig3-2-a}
		\end{minipage}%
	}
	\subfigure[]
	{
		\begin{minipage}[t]{0.5\linewidth}
			%\flushleft
			\centering
			\includegraphics[scale=0.55]{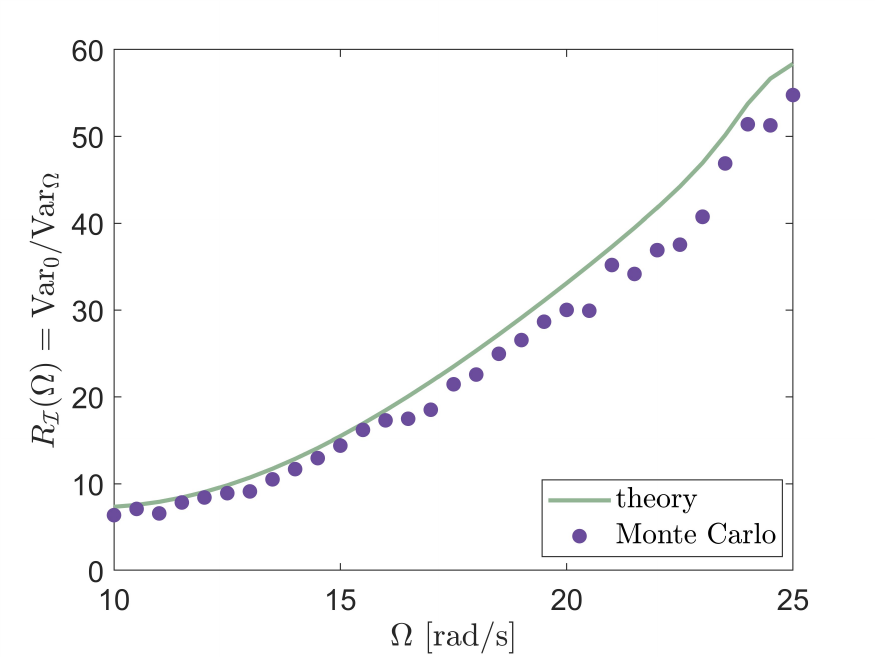}
			\label{fig3-2-b}
		\end{minipage}
	}
	\caption{$R_{\mathcal{I}}$ obtained from simulations and theory for the case $\omega_0\approx\omega_d$. (a) $C_0=1000$. (b) $C_0=1500$. The simulation results show good agreement with the theoretical predictions. When the transient response is treated as part of the estimation target, the Fisher information is significantly enhanced; even without energy normalization, the Fisher information ratio remains greater than unity. Moreover, the system’s initial conditions exert a pronounced influence on the Fisher information.}
	\label{fig3-2}
\end{figure}

Similarly, we performed numerical validation of $R_{\mathcal{I}}(\Omega)$ for the second case when $\omega_0\approx\omega_d$ using the same set of parameters. In this simulation, we set $\omega_0=\omega_d$ and assign the system two different initial conditions to examine the influence of the initial state on the Fisher information, $C_0 = 1000$ and $1500$. The results are shown in Fig.~\ref{fig3-2}; in both cases, the initial velocity is zero. A comparison between Figs.~\ref{fig3-2-a} and~\ref{fig3-2-b} shows that the theoretical curve predicted by the expression of $R_{\mathcal{I}}(\Omega)$ under the condition $\omega_0\approx\omega_d$ agrees well with the Monte Carlo results, thereby confirming its validity. Moreover, it is evident that the initial displacement has a pronounced effect on the Fisher information: a larger initial displacement leads to a more significant enhancement of the Fisher information. This observation further verifies the previously derived conclusions regarding the relationship between the system’s initial conditions and the Fisher information.

\begin{figure}[H]%fig3
	\makeatletter
	\renewcommand{\fnum@figure}{\textbf{Fig.} \thefigure.\@gobble}
	\makeatother
	\subfigure[]
	{
		\begin{minipage}[t]{0.5\linewidth}
			%\flushright
			\centering
			\includegraphics[scale=0.55]{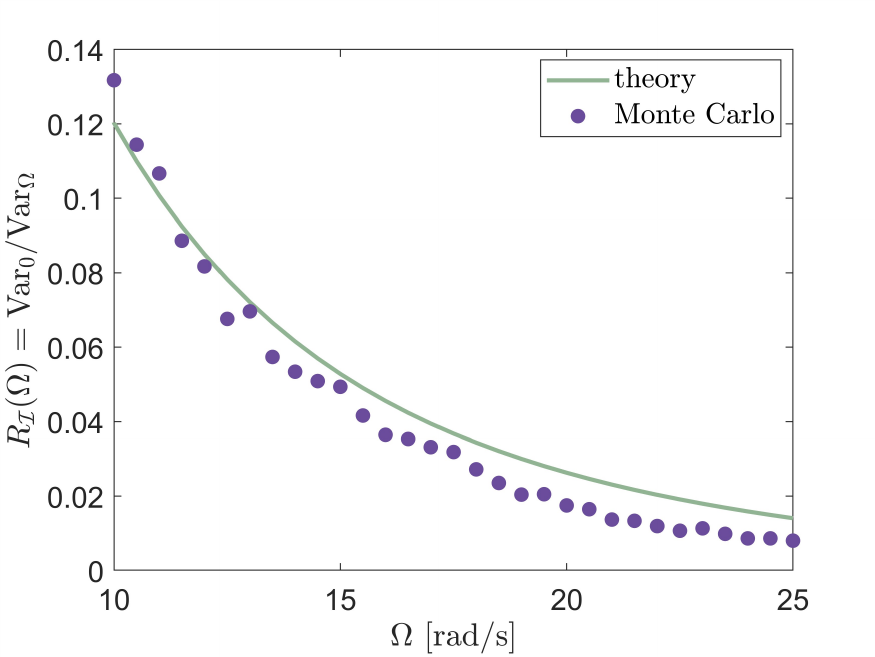}
			\label{fig3-3-a}
		\end{minipage}%
	}
	\subfigure[]
	{
		\begin{minipage}[t]{0.5\linewidth}
			%\flushleft
			\centering
			\includegraphics[scale=0.55]{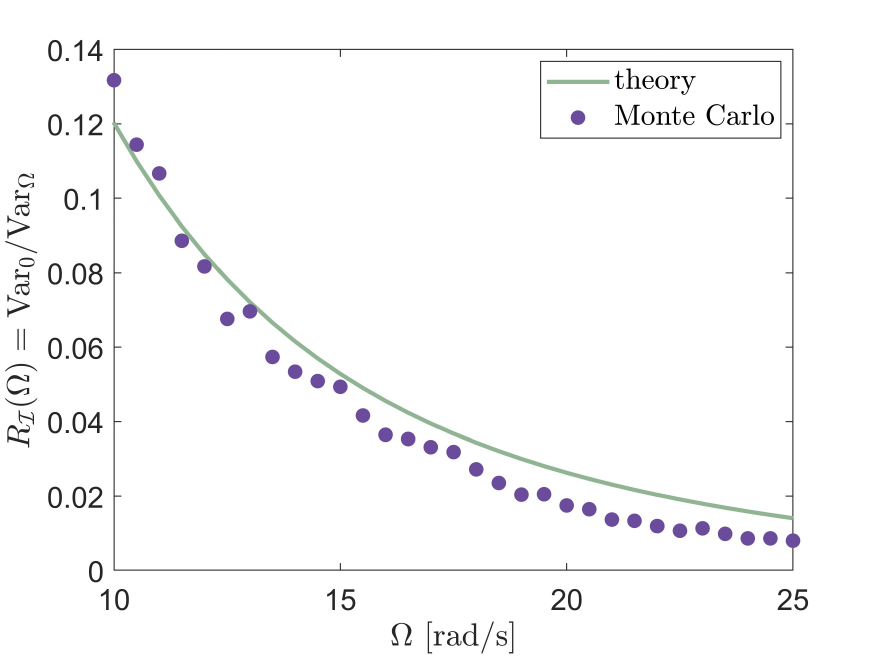}
			\label{fig3-3-b}
		\end{minipage}
	}
	\caption{$R_{\mathcal{I}}$ obtained from simulations and theory for the case $\omega_0\not \approx\omega_d$. (a) $C_0=50$. (b) $C_0=200$. The Fisher information of the system decreases monotonically with increasing $\Omega$ and is independent of the initial conditions. The deviation of points from the theoretical prediction is identical in both cases and it arises due to systematic errors caused by the specific estimation method. } 
	\label{fig3-3}
\end{figure}

For the third case $\omega_0 \not \approx \omega_d$, we set $\omega_0 =1.02 \omega_d$ and, while keeping all other parameters unchanged, apply two different initial displacements to the system, $C_0 = 50$ and $200$. The Monte Carlo results displayed in Fig.~\ref{fig3-3} show an excellent agreement with the theoretical curves, thereby verifying the correctness of $R_{\mathcal{I}}(\Omega)$ in this case. 
%\textbf{move to theory section:}
%When the transient response is treated as an unknown linear disturbance, the Fisher information decreases significantly, and the extent of this reduction is independent of the transient amplitude. It depends solely on the inner product between the basis vectors of the signal subspace spanning the transient response and that spanning the steady-state response. 
By comparing Fig.~\ref{fig3-3-a} and Fig.~\ref{fig3-3-b}, we observe that the amplitude of the transient response does not affect the Fisher information. Instead, the Fisher information depends on the inner product between the basis vectors spanning the subspace of  transient response signals and that of the steady-state response signals. In addition, it is worth noting that the discrepancy between the Monte Carlo results and the theoretical predictions remains nearly identical in the two simulations. These identical distributions indicate that the deviation does not arise from noise-induced random errors, but rather from systematic errors associated with the estimation method. Furthermore, a comparison between Fig.~\ref{fig3-2} and Fig.~\ref{fig3-3} indicates that the influence of the transient response on the Fisher information is highly significant.

\section{Demonstration with experimental data}
\label{Experimental validations}
In this section, an aluminum-alloy cantilever beam is used as the experimental specimen. We focus exclusively on the first flexural mode of the beam. Since the analysis is restricted to a narrow frequency band around the first resonance and the response in this range is dominated by the first flexural mode, the beam can be approximated as a single-degree-of-freedom system for the present purpose. We artificially add Gaussian white noise to the measured experimental data for our analysis. By comparing the variance of the peak-frequency estimates obtained under different excitations, we show that complex-frequency excitation yields more Fisher information and more accurate frequency estimation than normal excitation, and can even provide stable natural frequency estimates when the peak frequency under normal excitation is no longer identifiable.

\subsection{Experimental setup and procedure}
\begin{figure}[H]%fig1
	\makeatletter
	\renewcommand{\fnum@figure}{Fig. \thefigure.\@gobble}
	\makeatother
	\centering
	\includegraphics[scale=0.22]{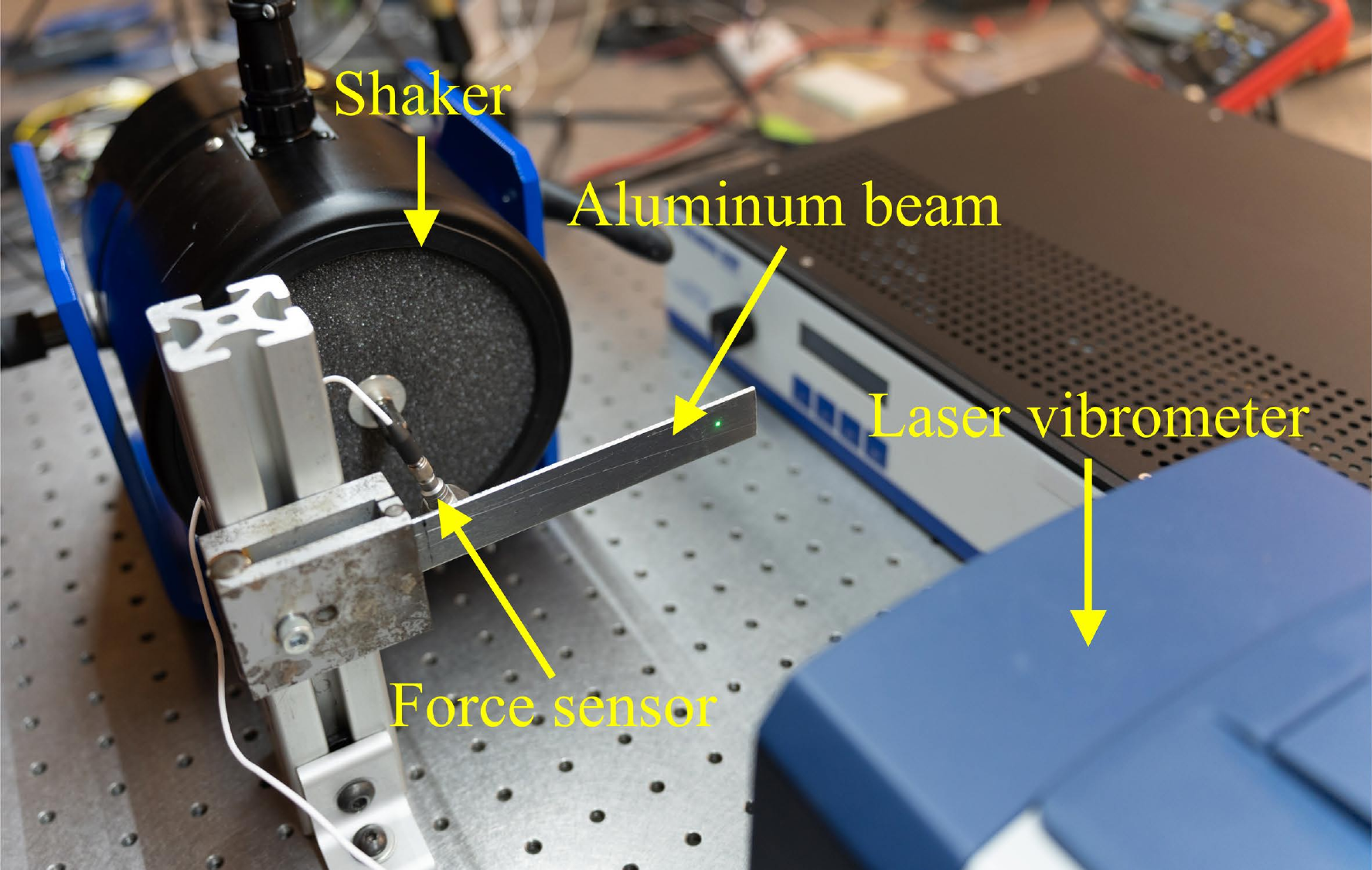}
	\caption{Experimental setup. A prescribed excitation is applied near the clamped end of an aluminum-alloy cantilever beam using a shaker, and the actual applied force is measured with a force sensor. A laser Doppler vibrometer is used to measure the out-of-plane velocity near the free end of the beam. From~\cite{li2025effective}.}
	\label{fig4-1}
\end{figure}

The experimental setup is shown in Fig.~\ref{fig4-1} \cite{li2025effective}. Excitation normal to the beam surface is applied near the clamped end of the cantilever beam using a Modal Shop 2025E modal shaker, so as to induce small-amplitude vibrations of the aluminum-alloy cantilever beam along the thickness direction. The actual applied excitation is measured by a force sensor mounted at the contact interface between the shaker and the beam. The vibration velocity at the free end of the beam is then measured and recorded using a PSV-500 Polytec scanning laser vibrometer (SLV) system for subsequent analysis.
We apply both a normal excitation and the corresponding complex-frequency excitation obtained through complex-frequency operator $T_{\Omega}$ to the beam, with the two excitations having the same initial amplitude. Gaussian white noise is then added to the measured displacement signal at the free end of the beam, and the resulting noisy displacement signal is used to calculate the amplitude-frequency response function $|H(\omega_0)|$ of the cantilever beam under the normal excitation and under the complex-frequency excitation framework, respectively. Because the system is lightly damped ($\zeta \approx 3.6 \times 10^{-3}$), the peak frequency $f_p$ of $|H(\omega_0)|$ can be taken as an approximation of the natural frequency.

In this experiment, the applied excitation is
\begin{equation*} \label{eq4-1}
F(t) = \begin{cases} 
F_0e^{-\Omega t}\sin\left(\frac{\omega_c}{N}t\right)\sin(\omega_ct), & 0 \le t \le {N\pi}/{\omega_c}, \\ 
0, & {N\pi}/{\omega_c} < t \le 9.9 \text{ s},
\end{cases}
\end{equation*}
where $F_0$ is the initial amplitude of the excitation, and its value is kept fixed throughout the experiment to ensure that the normal excitation and the complex-frequency excitation have the same initial amplitude. $\omega_c$ is the central frequency of the excitation. $N$ is a parameter that controls the width of the window function. When $\Omega > 0$, the amplitude of the excitation decays exponentially with $t$, and $F(t)$ therefore corresponds to a complex-frequency excitation. When $\Omega= 0$, the amplitude of $F(t)$ remains constant in time, reducing to a normal excitation.

\subsection{Case study}
We set $F_0=20$ N, $\omega_c=565$ rad/s, and $N=48$. Likewise, we set $\Omega=0$ and $1.47$ rad/s to generate the normal excitation and the complex-frequency excitation, respectively. Here, the value of $\Omega$ is chosen based on our approximate prior knowledge of $\omega_n$ and $\zeta$ ranges of the cantilever beam. For the measured vibration-velocity signal at the free end of the cantilever beam, discrete-time integration is performed to obtain the displacement signal, and additional strong Gaussian white noise with $\mathrm{SNR}=-25$ dB is then added to the displacement signal. For the force signal $F_n(t)$ and the noisy displacement signal $x_n(t)$ obtained under normal excitation, the amplitude-frequency response function $|H_n(\omega_0)|$ can be obtained directly by applying the fast Fourier transform (FFT). In contrast, for the force signal $F_c(t)$ and the noisy displacement signal $x_c(t)$ obtained under complex-frequency excitation, the mapping $T_m = T_{\Omega}^{-1}$ described in Sec. 2.1 is first applied to obtain the target excitation signal $f(t)$ and the noisy mapped displacement signal $\tilde{z}(t)$. An FFT is then performed on $f(t)$ and $\tilde{z}(t)$ to compute $|H_c(\omega_0)|$ of the cantilever beam under complex frequency excitation. 

\begin{figure}[H]%fig3
	\makeatletter
	\renewcommand{\fnum@figure}{\textbf{Fig.} \thefigure.\@gobble}
	\makeatother
	\subfigure[]
	{
		\begin{minipage}[t]{0.22\linewidth}
			%\flushright
			\centering
			\includegraphics[scale=0.52]{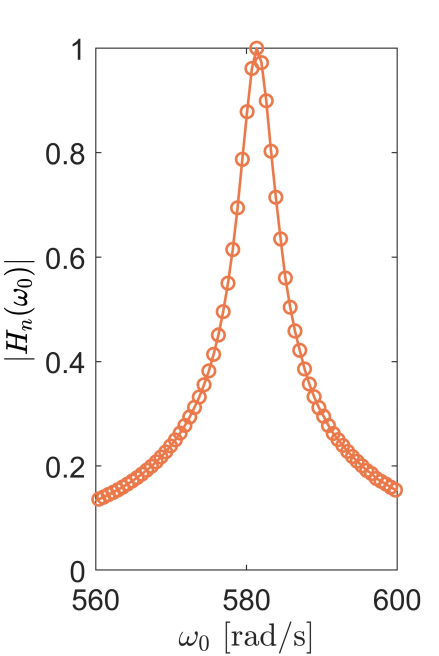}
			\label{fig4-2-a}
		\end{minipage}%
	}
	\subfigure[]
	{
		\begin{minipage}[t]{0.22\linewidth}
			%\flushleft
			\centering
			\includegraphics[scale=0.52]{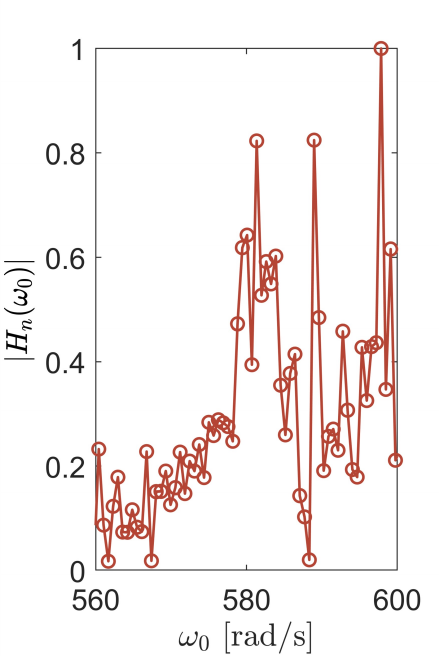}
			\label{fig4-2-b}
		\end{minipage}
	}
    \subfigure[]
	{
		\begin{minipage}[t]{0.22\linewidth}
			%\flushleft
			\centering
			\includegraphics[scale=0.52]{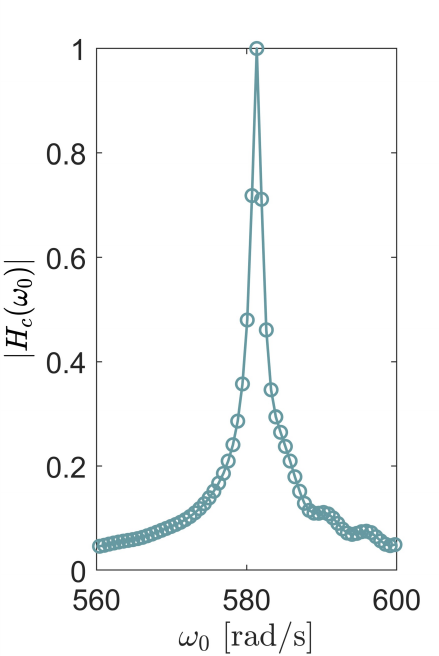}
			\label{fig4-2-c}
		\end{minipage}
	}
    \subfigure[]
	{
		\begin{minipage}[t]{0.22\linewidth}
			%\flushleft
			\centering
			\includegraphics[scale=0.52]{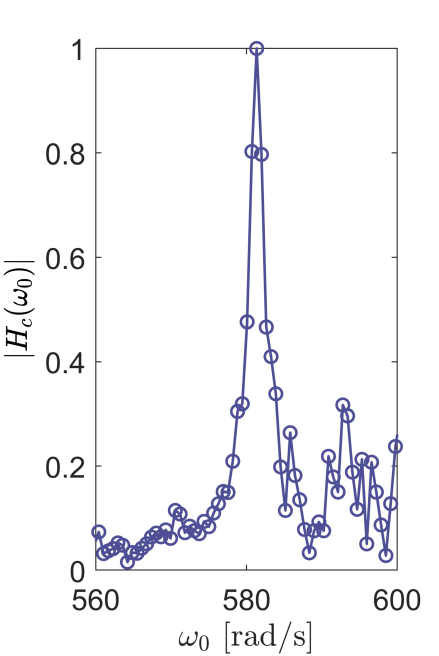}
			\label{fig4-2-d}
		\end{minipage}
	}
	\caption{Amplitude-frequency response functions under normal excitation and complex-frequency excitation. Normal excitation (a) without and (b) with additional noise. Complex-frequency excitation (c) without and (d) with additional noise. Under strong Gaussian white noise, the resonance peak in $|H_n(\omega_0)|$ under normal excitation is no longer identifiable. In contrast, for $|H_c(\omega_0)|$ under complex-frequency excitation with noise of the same $\mathrm{SNR}$, the resonance peak remains well preserved, and $f_p$ remains unchanged.}
	\label{fig4-2}
\end{figure}

The resulting amplitude-frequency response functions are shown in Fig. \ref{fig4-2}. For clarity, each curve is normalized separately. The amplitude-frequency response function $|H_n(\omega_0)|$ of the first flexural mode of the aluminum-alloy cantilever beam under normal excitation based on the actual measured data is shown in Fig. \ref{fig4-2-a}. After adding the above noise to the displacement signal $x_n(t)$, the resulting $|H_n(\omega_0)|$ is shown in Fig. \ref{fig4-2-b}. We observe that under normal excitation, the introduction of noise renders the resonance peak in $|H_n(\omega_0)|$ no longer identifiable, and the peak frequency $f_p$ can no longer be estimated correctly. For the complex-frequency excitation framework, the $|H_c(\omega_0)|$ from the mapped data is shown in Fig. \ref{fig4-2-c}. We then add Gaussian white noise with the same $\mathrm{SNR}$ to the displacement signal $x_c(t)$, and the resulting $|H_c(\omega_0)|$ under strong-noise conditions after mapping $T_{\Omega}^{-1}$ is shown in Fig. \ref{fig4-2-d}. A comparison of Fig. \ref{fig4-2-b} and Fig. \ref{fig4-2-d} demonstrates that under the same noise level, the resonance peak in $|H_c(\omega_0)|$ under complex-frequency excitation remains correctly identifiable. Such excitations yield more Fisher information over conventional harmonic excitations, which gives it a significant advantage in frequency estimation. In addition, by comparing Fig. \ref{fig4-2-a}, Fig. \ref{fig4-2-c}, and Fig. \ref{fig4-2-d}, we observe that the peak frequency $f_p$ obtained within the complex-frequency excitation framework remains accurate even under strong-noise conditions.

To further verify the advantage of the complex-frequency excitation framework in frequency estimation, we keep $\mathrm{SNR}=-25$ dB and add ten different realizations of random Gaussian white noise to the displacement signals obtained under normal excitation and complex-frequency excitation, respectively. The corresponding amplitude-frequency response functions are then calculated using the procedure described above. We take the frequency corresponding to the peak of amplitude-frequency response functions within the range 560 rad/s to 600 rad/s as the peak frequency $f_p$, and compute the variance of $f_p$. The results listed in Table \ref{tab2} show that the variance of $f_p$ under complex-frequency excitation is 0.072, which is far smaller than the corresponding value of 63.32 under normal excitation. In addition, we take the peak frequency under the noise-free condition in Fig. \ref{fig4-2-a} as the true value $f_{p,\mathrm{true}} = 581.353~\mathrm{rad/s}$. The absolute error of $f_p$ under complex-frequency excitation in the strong-noise case is then 0.127 rad/s, which is similarly significantly smaller than the corresponding value of 6.918 rad/s under normal excitation. These results demonstrate that, under strong-noise conditions, the Fisher-information gain provided by the proposed framework makes the natural frequency estimation markedly more stable and accurate. Moreover, this Fisher-information gain extends the operating limit of frequency estimation in strong noise, allowing resonance peaks that are no longer identifiable under normal excitation to be successfully recovered.

\begin{table}[H]
\centering
\caption{Peak-frequency estimates $f_p$ under strong-noise conditions ($\mathrm{SNR}=-25~\mathrm{dB}$) for ten noise realizations under normal excitation and complex-frequency excitation. The true value is $f_{p,\mathrm{true}} = 581.353~\mathrm{rad/s}$. The absolute error is defined as the absolute difference between the mean value of the ten estimated peak frequencies and $f_{p,\mathrm{true}}$.}
\label{tab2}
\begin{tabular}{c
                S[table-format=3.3]
                S[table-format=3.3]}
\toprule
\multirow{2}{*}{Realization} &
\multicolumn{1}{c}{$f_p$ under normal excitation} &
\multicolumn{1}{c}{$f_p$ under complex-frequency excitation} \\
&
\multicolumn{1}{c}{(rad/s)} &
\multicolumn{1}{c}{(rad/s)} \\
\midrule
1  & 582.623 & 581.353 \\
2  & 598.489 & 581.353 \\
3  & 595.951 & 581.353 \\
4  & 599.124 & 581.353 \\
5  & 581.353 & 581.353 \\
6  & 581.988 & 581.988 \\
7  & 580.719 & 581.353 \\
8  & 581.988 & 581.988 \\
9  & 595.951 & 581.353 \\
10 & 584.527 & 581.353 \\
\midrule
Mean value     & 588.271 & 581.480 \\
Absolute error & 6.918   & 0.127   \\
Variance       & 63.325  & 0.072   \\
\bottomrule
\end{tabular}
\end{table}

\section{Conclusions}
\label{Conclusions}

We propose a complex-frequency-excitation-based method for estimating system natural frequencies under strong-noise conditions, and compare its accuracy and robustness to conventional approaches.
%In addition, we investigate the Fisher information of a second-order underdamped linear time-invariant system with Gaussian noise under complex-frequency excitation\textbf{too long, split into 2 sentences}, as well as the corresponding Fisher information ratio relative to classical harmonic excitation. 
To establish the superior performance of the proposed method, we investigate the Fisher information under the complex-frequency framework, as well as the corresponding information ratio to the normal excitation. The results translate to any second-order linear time-invariant system.

Our theoretical results indicate that it is the exponential variation of the excitation amplitude, rather than the mapping $T_m$ applied from the actual output to the mapped output, that governs the Fisher information of the system. In particular, the excitation duration and the amplitude-decay factor are the two most critical parameters governing the Fisher information, as they determine the temporal distribution of the complex-frequency excitation amplitude. When the oscillation frequency of the complex-frequency excitation matches the damped natural frequency of the system, the transient response of the system  contributes to the mapped output, leading to a sharp increase in the Fisher information. 

Moreover, the Fisher information under complex-frequency excitation depends not only on the excitation parameters but also on the initial state of the system; the initial energy of the system can yield a substantial enhancement in the Fisher information. This observation provides new insight into the noise robustness of complex-frequency excitation. We verify this insight regarding the initial condition under complex-frequency excitation with numerical simulations. On an experimental dataset, the resonance peak of the cantilever beam that is obscured by noise under harmonic excitations, is recovered successfully by using complex-frequency excitation. This comparison demonstrates the significant advantages of our  complex-frequency framework on the accuracy and robustness under the strong noise compared to the conventional harmonic excitation.
Here, we assumed that the natural frequency, damping and thus the optimal amplitude-decay factor are approximately known in the structure. Future work will address how to determine an appropriate $\Omega$ value when the system parameters are unknown. 

This work not only enables accurate and robust estimation of system natural frequencies under strong-noise conditions, but also deepens our understanding of the fundamental nature of complex-frequency excitations. The results provide guidance for the design of complex-frequency excitations in optics, sensing, and elastodynamics for structural health monitoring and material characterization applications. 

\section*{Acknowledgements}
WL and RKP gratefully acknowledge support from NSF grant 2503599.

\section*{Appendix A: General expression for $B_{\Omega}$}
We provide a detailed derivation of the amplitude $B_{\Omega}$ in Eq.~\eqref{eq1-13}. 
We set f(t) as the analytic-signal representation of a windowed tone burst \cite{reilly2002analytic}:
\begin{equation}\label{app1-1}
  f(t)=A\,w(t)\,e^{j\omega_0 t},\qquad 0\le t\le t_{final},
\end{equation}
and let the effective impulse response of the system under complex-frequency framework be
\begin{equation}
  h_\Omega(t)=\frac{\omega_n}{\omega_d}\,e^{-(\omega_n\zeta-\Omega)t}\,\sin(\omega_d t)\,u(t),
\end{equation}
where $u(t)$ is the unit step.

Factoring out the fast carrier $e^{j\omega_0 t}$, we define the complex envelope $G(\omega_0;\Omega,t)$ of the output as
\begin{equation}
  G(\omega_0;\Omega,t)\triangleq\int_{0}^{t} h_\Omega(\tau)\,w(t-\tau)\,e^{-j\omega_0 \tau}\,d\tau .
\end{equation}
Then the noise-free mapped output can be expressed as 
\begin{equation}
  z_s(t)=\Re\!\big\{A\,e^{j\omega_0 t}\,G(\omega_0;\Omega,t)\big\}.
\end{equation}
Therefore, there is
\begin{equation}
  B_\Omega(t)\triangleq A\,\big|G(\omega_0;\Omega,t)\big|.
\end{equation}
If $w(t)$ varies slowly, $B_\Omega(t)$ changes only mildly within the interior of the burst and can be regarded as approximately constant; we then write $B_\Omega(t)\approx B_\Omega$
and use the narrowband model:
\begin{equation}
  z_s(t)\approx B_\Omega\,w(t)\cos(\omega_0 t+\psi)
\;+\; C\,e^{(\Omega-\zeta\omega_n)t}\cos(\omega_d t+\psi_c).
\end{equation}
The above expressions represent generalized forms of $B_{\Omega}$ and are not restricted to the specific case of $\Omega = \omega_n \zeta$ and $\omega_0 \approx \omega_d$; all other conditions remain valid. For convenience, however, in other cases we still define $B_{\Omega} = A|H(j\omega_0 - \Omega)|$.

\section*{Appendix B: Derivation of the Fisher information when $\omega_0 \not \approx \omega_d$}
Based on Eqs.~\eqref{eq1-51}-\eqref{eq1-53}, we combine $\omega_0$, $\alpha$, and $\beta$ into the following vector:
\begin{equation} \label{eq1-54}
	\bm{\theta}=(\omega_0,\,\mathbf{u}),
\end{equation}
\begin{equation} \label{eq1-55}
	\mathbf{u}=\begin{bmatrix}
 \alpha & \beta
\end{bmatrix}^{\top}.
\end{equation}
Then, the mean $\mu(t;\bm{\theta})$ of the observed mapped output can be written as
\begin{equation} \label{eq1-55}
	\mu(t;\bm{\theta})=z_{ss}(t;\omega_0)+K(t)\mathbf{u},
\end{equation}
\begin{equation} \label{eq1-56}
	K(t)=\begin{bmatrix}
 e^{(\Omega-\omega_n\zeta)t}\cos(\omega_dt) & e^{(\Omega-\omega_n\zeta)t}\sin(\omega_dt) 
\end{bmatrix}.
\end{equation}
After discretization according to the sampling process, we obtain
\begin{equation} \label{eq1-57}
	\mathbf{z}\sim\mathcal{N}\left ( \bm{\mu}(\bm{\theta}),\mathbf{\Sigma}_\Omega \right ).
\end{equation}
We define
\begin{equation} \label{eq1-58}
	\mathbf{g}\triangleq \partial_{\omega_0}\bm{\mu}=\partial_{\omega_0}\mathbf{z}_{ss},
\end{equation}
\begin{equation} \label{eq1-59}
	\mathbf{K}\triangleq \frac{\partial\bm{\mu}}{\partial \mathbf{u}}=\begin{bmatrix}
 \mathbf{k}_1 & \mathbf{k}_2
\end{bmatrix},
\end{equation}
where $\mathbf{k}_1$ and $\mathbf{k}_2$ are the discrete sampling vectors of $e^{(\Omega-\omega_n \zeta) t} \cos(\omega_d t)$ and $e^{(\Omega-\omega_n\zeta)t}\sin(\omega_dt) $, respectively.

For $\mathcal{N}\left ( \bm{\mu}(\bm{\theta}),\mathbf{\Sigma}_\Omega \right )$, the corresponding Fisher information can be written as
\begin{equation} \label{eq1-60}
	\bm{\mathcal{I}_\theta}=\left ( \partial_{\bm{\theta}} \bm{\mu}\right )^{\top}\Sigma_{\Omega}^{-1}\left ( \partial_{\bm{\theta}} \bm{\mu}\right ).
\end{equation}
Eq.~\eqref{eq1-60} can be rewritten as
\begin{equation} \label{eq1-61}
	\bm{\mathcal{I}_\theta}=\begin{bmatrix}
\mathcal{I}_{\omega_0 \omega_0}  & \mathcal{I}_{\omega_0 u} \\
\mathcal{I}_{u \omega_0}  & \mathcal{I}_{u u}
\end{bmatrix}=\begin{bmatrix}
\mathbf{g}^{\top}\bm{\Sigma}^{-1}_{\Omega}\mathbf{g} & \mathbf{g}^{\top}\bm{\Sigma}^{-1}_{\Omega}\mathbf{K} \\
\mathbf{K}^{\top}\bm{\Sigma}^{-1}_{\Omega}\mathbf{g}  & \mathbf{K}^{\top}\bm{\Sigma}^{-1}_{\Omega}\mathbf{K}
\end{bmatrix}.
\end{equation}
Because the parameter $\mathbf{u}$ is treated as an unknown linear disturbance, the Fisher information for $\omega_0$ can be calculated as
\begin{equation} \label{eq1-62}
	\mathcal{I}_{\Omega}=\mathcal{I}_{\omega_0 \omega_0}-\mathcal{I}_{\omega_0 u}\mathcal{I}_{u u}^{-1}\mathcal{I}_{u \omega_0}=\mathbf{g}^{\top}\bm{\Sigma}^{-1}_{\Omega}\mathbf{g} -\mathbf{g}^{\top}\bm{\Sigma}^{-1}_{\Omega}\mathbf{K}\left ( \mathbf{K}^{\top}\bm{\Sigma}^{-1}_{\Omega}\mathbf{K} \right )^{-1}\mathbf{K}^{\top}\bm{\Sigma}^{-1}_{\Omega}\mathbf{g} .
\end{equation}

For notational convenience, we introduce the following $\Omega$-dependent weighted inner product:
\begin{equation} \label{eq1-63}
	\left \langle f_1,f_2 \right \rangle_{\Omega}=\frac{1}{\sigma_m^2}\int_0^{t_{final}}e^{-2\Omega t}f_1(t)f_2(t)\mathrm{d}t,
\end{equation}
where $f_1(t)$ and $f_2(t)$ are the arbitrary functions of $t$. When sampling interval $\Delta t\to 0$, there are
\begin{equation} \label{eq1-64}
	\mathbf{g}^{\top}\bm{\Sigma}^{-1}_{\Omega}\mathbf{g}\to\left \langle g,g \right \rangle_\Omega,
\end{equation}
\begin{equation} \label{eq1-65}
	\mathbf{K}^{\top}\bm{\Sigma}^{-1}_{\Omega}\mathbf{K}\to \left [\left \langle k_i,k_j \right \rangle_\Omega \right ]^2_{i,\,j=1,\,2},
\end{equation}
\begin{equation} \label{eq1-66}
	\mathbf{K}^{\top}\bm{\Sigma}^{-1}_{\Omega}\mathbf{g}\to \left [\left \langle k_i,g \right \rangle_\Omega \right ]^2_{i,\,j=1,\,2},
\end{equation}
Based on Eq.~\eqref{eq1-52} and~\eqref{eq1-58}, we can obtain
\begin{equation} \label{eq1-67}
	g(t)\triangleq -tB_{\Omega}w(t)\sin(\omega_0t+\psi),
\end{equation}
Then, there is
\begin{equation} \label{eq1-68}
	\left \langle g,g \right \rangle_\Omega=\frac{B_{\Omega}^2}{\sigma_m^2}\int_0^{t_{final}}e^{-2\Omega t}t^2w^2(t)\sin^2(\omega_0t+\psi)\mathrm{d}t\approx \frac{B_{\Omega}^2}{2\sigma_m^2}J_{ss}(\Omega).
\end{equation}
Therefore, $\mathbf{g}^{\top}\bm{\Sigma}^{-1}_{\Omega}\mathbf{g}$ can be calculated as
\begin{equation} \label{eq1-69}
	\mathbf{g}^{\top}\bm{\Sigma}^{-1}_{\Omega}\mathbf{g}\approx \frac{B_{\Omega}^2}{2\sigma_m^2}J_{ss}(\Omega).
\end{equation}

Similarly, we can get
\begin{equation} \label{eq1-70}
	\left \langle k_i,k_j \right \rangle_\Omega=\frac{1}{\sigma_m^2}\int_0^{t_{final}}e^{-2\omega_n\zeta t}\phi_i(t)\phi_j(t)\mathrm{d}t,
\end{equation}
where $\phi_1(t)=\cos(\omega_d t)$ and $\phi_2(t)=\sin(\omega_d t)$. Then, there are
\begin{equation} \label{eq1-71}
	\left \langle k_1,k_1 \right \rangle_\Omega\approx\left \langle k_2,k_2 \right \rangle_\Omega\approx \frac{1}{2\sigma_m^2}\int_0^{t_{final}}e^{-2\omega_n\zeta t}\mathrm{d}t,
\end{equation}
\begin{equation} \label{eq1-72}
	\left \langle k_1,k_2 \right \rangle_\Omega\approx\left \langle k_2,k_1 \right \rangle_\Omega\approx 0.
\end{equation}
And we define
\begin{equation} \label{eq1-73}
	L_{tr}\triangleq\int_0^{t_{final}}e^{-2\omega_n\zeta t}\mathrm{d}t.
\end{equation}
Therefore, we can get
\begin{equation} \label{eq1-74}
	\mathbf{K}^{\top}\bm{\Sigma}^{-1}_{\Omega}\mathbf{K}\approx \frac{L_{tr}}{2\sigma^2_m}\mathbf{I}.
\end{equation}

For the mixed term $\left \langle k_i,g \right \rangle_\Omega$, let's take $k_1$ as an example:
\begin{equation} \label{eq1-75}
	\left \langle k_1,g \right \rangle_\Omega=-\frac{B_{\Omega}}{\sigma_m^2}\int_0^{t_{final}}te^{-(\Omega+\omega_n\zeta) t}w(t)\cos(\omega_dt)\sin(\omega_0t+\psi)\mathrm{d}t.
\end{equation}
% Since $\cos(\omega_d t)\sin(\omega_0t+)$
And we define
\begin{equation} \label{eq1-76}
	L_{mix}(\Omega)\triangleq \int_0^{t_{final}}te^{-(\Omega+\omega_n\zeta) t}w(t)\mathrm{d}t.
\end{equation}
Then, we can obtain
\begin{equation} \label{eq1-77}
	\left \langle k_1,g \right \rangle_\Omega=-\frac{B_{\Omega}}{\sigma_m^2}L_{mix}(\Omega)r_1(\Omega),
\end{equation}
\begin{equation} \label{eq1-78}
	r_1(\Omega)=\frac{\int_0^{t_{final}}te^{-(\Omega+\omega_n\zeta) t}w(t)\cos(\omega_dt)\sin(\omega_0t+\psi)\mathrm{d}t}{L_{mix}(\Omega)}.
\end{equation}
Similarly, for $\left \langle k_2,g \right \rangle_\Omega$, there are
\begin{equation} \label{eq1-79}
	\left \langle k_2,g \right \rangle_\Omega=-\frac{B_{\Omega}}{\sigma_m^2}L_{mix}(\Omega)r_2(\Omega),
\end{equation}
\begin{equation} \label{eq1-80}
	r_2(\Omega)=\frac{\int_0^{t_{final}}te^{-(\Omega+\omega_n\zeta) t}w(t)\sin(\omega_dt)\sin(\omega_0t+\psi)\mathrm{d}t}{L_{mix}(\Omega)}.
\end{equation}
$r_1(\Omega)$ and $r_2(\Omega)$ are the weighted correlation coefficients between $\sin(\omega_0 t + \psi)$ and the transient basis functions, and they satisfy $0\le r_1^2(\Omega)+r_2^2(\Omega)\le 1$. The proof is omitted here.

Eq.~\eqref{eq1-62} can be rewritten as
\begin{align}\label{eq1-81}
	\mathcal{I}_{\Omega}&=\mathbf{g}^{\top}\bm{\Sigma}^{-1}_{\Omega}\mathbf{g}-\left (\mathbf{K}^{\top}\bm{\Sigma}^{-1}_{\Omega}\mathbf{g}\right )^{\top}\left ( \mathbf{K}^{\top}\bm{\Sigma}^{-1}_{\Omega}\mathbf{K} \right )^{-1}\left (\mathbf{K}^{\top}\bm{\Sigma}^{-1}_{\Omega}\mathbf{g}\right ).
\end{align}
Therefore, there is
\begin{equation} \label{eq1-82}
	\mathcal{I}_{\Omega}\approx\frac{B_{\Omega}^2}{2\sigma_m^2}\left[ J_{ss}(\Omega)-\frac{4L_{mix}^2(\Omega)}{L_{tr}}\left( r_1^2(\Omega)+r_2^2(\Omega)\right) \right].
\end{equation}

\section*{Appendix C: Derivation of the information gain under colored noise}
We assume that the colored noise in the actual output $x(t)$ is a zero-mean stationary colored Gaussian noise, whose autocovariance is $r(\tau)$ and PSD is $S_m(\omega)$. The discrete noise vector $\mathbf{m}$ can be written as
\begin{equation} \label{eq1-19}
	\mathbf{m}=[m_0,m_1,...,m_{N-1}]^{\top}.
\end{equation}
where $N$ is the number of sampling points. At this point, the covariance matrix $\mathbf{R}$ of the noise can be calculated as
\begin{equation} \label{eq1-20}
	\mathbf{R}=\mathrm{E}[\mathbf{mm}^{\top}],
\end{equation}
\begin{equation} \label{eq1-21}
	R_{ij}=r\left ((i-j)/f_s \right ),
\end{equation}
where $f_s$ is the sampling frequency. %The covariance $\Sigma_{\Omega}$ of the colored noise $\mathbf{n}_c=\mathbf{Dm}$ of the target output can be calculated as 
The colored noise vector $\mathbf{n}_c$ of the mapped output can be expressed as
\begin{equation} \label{eq1-22}
	\mathbf{n}_c=\mathbf{Dm},
\end{equation}
\begin{equation} \label{eq1-23}
	\mathbf{D}=\mathrm{diag}(e^{\Omega t_0},e^{\Omega t_1},...,e^{\Omega t_{N-1}}).
\end{equation}
Then, the covariance $\Sigma_{\Omega}$ of $\mathbf{n}_c$ can be calculated as
\begin{equation} \label{eq1-24}
	\Sigma_{\Omega}=\mathrm{E}[\mathbf{nn}^{\top}]=\mathbf{DRD}.
\end{equation}

At this point, the mapped output model can be written as
\begin{equation} \label{eq1-25}
	\mathbf{z}=\mathbf{z}_s+\mathbf{n}_c.
\end{equation}
Similarly, the Fisher information matrix can be calculated as
\begin{equation} \label{eq1-26}
	\mathcal{I}_{\Omega}(\omega_0)=(\partial_{\omega_0}\mathbf{z}_s)^\top\Sigma_{\Omega}^{-1}(\partial_{\omega_0}\mathbf{z}_s).
\end{equation}
Since $\Sigma_{\Omega}^{-1}=(\mathbf{DRD})^{-1}=\mathbf{D}^{-1}\mathbf{R}^{-1}\mathbf{D}^{-1}$ and $\mathbf{x}_s\triangleq\mathbf{D}^{-1}\mathbf{z}_s$, Eq.~\eqref{eq1-26} can be rewritten as
\begin{equation} \label{eq1-27}
	\mathcal{I}_{\Omega}(\omega_0)=(\partial_{\omega_0}\mathbf{x}_s)^\top\mathbf{R}^{-1}(\partial_{\omega_0}\mathbf{x}_s).
\end{equation}

Since $\mathbf{R}$ is a Toeplitz matrix, when the time window length is sufficiently long, it can be approximated as circulant and decomposed spectrally as
\begin{equation} \label{eq1-28}
	\mathbf{R}=\mathbf{F}^\mathrm{H}\mathbf{\Lambda F},
\end{equation}
where $\mathbf{F}$ is the discrete Fourier transform (DFT) matrix, $\mathbf{F}^\mathrm{H}$ is its Hermitian transpose, and $\mathbf{\Lambda}=\mathrm{diag}(S_m(\omega_k))$. Let us take the whitening operator $\mathbf{Q}=\mathbf{\Lambda}^{-\frac{1}{2}}\mathbf{F}$; then the whitened ideal actual output $\tilde{\mathbf{x}}$ can be expressed as
\begin{equation} \label{eq1-29}
	\tilde{\mathbf{x}}=\mathbf{Q}\mathbf{x}_s.
\end{equation}
Then, there are
\begin{equation} \label{eq1-30}
	\partial_{\omega_0}\tilde{\mathbf{x}}=\mathbf{\Lambda}^{-\frac{1}{2}}\mathbf{F}\partial_{\omega_0}\mathbf{x}_s,
\end{equation}
\begin{equation} \label{eq1-31}
\begin{split}
	||\partial_{\omega_0}\tilde{\mathbf{x}}||^2_2 
	&= (\partial_{\omega_0}\mathbf{x}_s)^{\mathrm{H}}\mathbf{F}^{\mathrm{H}}\mathbf{\Lambda}^{-1}\mathbf{F}(\partial_{\omega_0}\mathbf{x}_s) \\
    &=
    \sum_{k=0}^{N-1} \frac{|\mathbf{F}\partial _{\omega_0}\mathbf{x}_s  |^2}{S_m(\omega_k)} \\
    &=
    \sum_{k=0}^{N-1} \frac{|\partial _{\omega_0}X_{\Omega} (\omega_k) |^2}{S_m(\omega_k)}.
\end{split}
\end{equation}
When $N$ is sufficiently large, using the Riemann sum to approximate the integral, Eq.~\eqref{eq1-31} can be rewritten as
\begin{equation} \label{eq1-32}
	||\partial_{\omega_0}\tilde{\mathbf{x}}||^2_2=\frac{1}{2\pi}\int_{-\pi f_s}^{\pi f_s}\frac{|\partial _{\omega_0}X_{\Omega} (j\omega) |^2}{S_m(\omega)}\mathrm{d}\omega.
\end{equation}

Let the window function $w(t)$ vary slowly and the signal be narrowband; then $\partial_{\omega_0} X_{\Omega}(j\omega)$ is concentrated only within $\Xi=[\omega_0-B,\omega_0+B]\cup[-\omega_0-B,-\omega_0+B]$, where $B$ is the bandwidth and $B\ll \omega_0$. Assume there exists a $\epsilon(\omega)$ such that $\epsilon_{\max}\ll1$, so that
\begin{equation} \label{eq1-33}
	S_m(\omega)=S_m(\omega_0)(1+\epsilon(\omega)), \ \ \ \ |\epsilon(\omega)|\leq \epsilon_{\max}.
\end{equation}
Then, based on Eq.~\eqref{eq1-32}, there is
\begin{equation} \label{eq1-34}
	\frac{1}{2\pi}\frac{1}{{S_m(\omega_0)(1+\epsilon_{\max})}}\int_{\Xi}|\partial _{\omega_0}X_{\Omega} |^2\mathrm{d}\omega\ \leq \ \mathcal{I}_{\Omega}\ \leq \ \frac{1}{2\pi}\frac{1}{{S_m(\omega_0)(1-\epsilon_{\max})}}\int_{\Xi}|\partial _{\omega_0}X_{\Omega} |^2\mathrm{d}\omega.
\end{equation}
We divide Eq.~\eqref{eq1-32} into two parts, the in-band and out-of-band regions:
\begin{equation} \label{eq1-35}
	\mathcal{I}_{\Omega}=\frac{1}{2\pi}\left ( \int_{\Xi} \frac{|\partial_{\omega_0}X_{\Omega}|^2}{S_m}\mathrm{d}\omega + \int_{\Xi^{c}} \frac{|\partial_{\omega_0}X_{\Omega}|^2}{S_m}\mathrm{d}\omega \right ).
\end{equation}
Then we can obtain
\begin{equation} \label{eq1-36}
	\mathcal{I}_{\Omega}=\frac{\int_{\Xi} |\partial_{\omega_0}X_{\Omega}|^2\mathrm{d}\omega}{2\pi S_m(\omega_0)}\cdot\left ( 1+\mathcal{O}(\epsilon_{\max}) + \mathcal{O}(\eta) \right ),
\end{equation}
\begin{equation} \label{eq1-37}
	\eta = \frac{\int_{\Xi^{c}} |\partial_{\omega_0}X_{\Omega}|^2\mathrm{d}\omega}{\int_{\Xi} |\partial_{\omega_0}X_{\Omega}|^2\mathrm{d}\omega} \ll 1.
\end{equation}
Therefore, $\mathcal{I}_{\Omega}$ can be simplified as
\begin{equation} \label{eq1-38}
	\mathcal{I}_{\Omega} \approx \frac{1}{2\pi{S_m(\omega_0)}}\int_{\Xi}|\partial _{\omega_0}X_{\Omega} |^2\mathrm{d}\omega.
\end{equation}
From Eq. (38), it can be seen that the Fisher information under colored noise is $(\sigma_n)^2/{S_m(\omega_0)}$ times that under white noise. The Fisher information ratio is calculated as
\begin{equation} \label{eq1-39}
	R_{\mathcal{I}}(\Omega)=\frac{\mathcal{I}_{\Omega}}{\mathcal{I}_{0}}\approx\left ( \frac{B_{\Omega}}{B_0} \right ) ^2\frac{J_{ss}(\Omega)}{J_{ss}(0)}=\left ( \frac{|H(s-\Omega)|}{|H(s)|} \right ) ^2\frac{J_{ss}(\Omega)}{J_{ss}(0)}.
\end{equation}
Moreover, when the PSD of the colored noise, $S_m(\omega)$, fluctuates significantly within $\Xi$ and thus cannot be treated as constant, the Fisher information $\mathcal{I}_{\Omega}$ is weighted in the frequency domain with $1/S_m(\omega)$ as the weighting factor:
\begin{equation} \label{eq1-40}
	\mathcal{I}_{\Omega}=\frac{1}{2\pi}\int_{-\infty}^{\infty}\frac{|\partial _{\omega_0}X_{\Omega} (j\omega) |^2}{S_m(\omega)}\mathrm{d}\omega.
\end{equation}
We can also consider only the influence of noise within the frequency band $\Xi$. In this case, the calculation of $\mathcal{I}_{\Omega}$ can be simplified as
\begin{equation} \label{eq1-41}
	\mathcal{I}_{\Omega}\approx\frac{1}{2\pi}\int_{\Xi}\frac{|\partial _{\omega_0}X_{\Omega} (j\omega) |^2}{S_m(\omega)}\mathrm{d}\omega=\frac{1}{2\pi}\int_{\Xi}\frac{|\partial _{\omega_0}H (j\omega-\Omega) f(j\omega) |^2}{S_m(\omega)}\mathrm{d}\omega.
\end{equation}

Similarly, when $\Omega$ is no longer assumed to be much smaller than $\omega_n \zeta$, the expressions for the Fisher information ratio are given by Eq.~\eqref{eq1-50} and~\eqref{eq1-83}; the derivations are omitted here. In addition, the energy normalization introduced in Sec.~\ref{Theory} remains applicable to the information ratio under colored-noise conditions.

\bibliography{bibref}
\bibliographystyle{unsrt}
\end{document}